\documentclass{aa}
\usepackage{natbib}
\usepackage{graphicx}
\usepackage{graphicx}
\usepackage{savesym}
\usepackage{amsmath}
\savesymbol{iint}
\usepackage{txfonts}
\restoresymbol{TXF}{iint}
\usepackage{amssymb}
\usepackage{epsfig}
\usepackage{multirow}
\usepackage{rotating}
\newcommand{\xmm}{\emph{XMM-Newton}}
\newcommand{\chandra}{\emph{Chandra}}

\newcommand{\lumunit}{erg~s$^{-1}$}
\newcommand{\fluxunit}{erg~s$^{-1}$~cm$^{-2}$}
\newcommand{\avg}[1]{\left< #1 \right>}
\usepackage{hyperref}

\begin{document}
\title{The \xmm~Wide Angle Survey (XWAS)}

\author{P. Esquej\inst{1,2,3,4},
M. Page\inst{5},
F. J. Carrera\inst{3},
S. Mateos\inst{3},
J. Tedds\inst{2}, 
M. G. Watson\inst{2},
A. Corral\inst{6},
J. Ebrero\inst{7},
M. Krumpe\inst{8,9,10},
S. R. Rosen\inst{2},
M. T. Ceballos\inst{3},
A. Schwope\inst{10}, 
C. Page\inst{2},
A. Alonso-Herrero\inst{3}\thanks{Augusto Gonz\'alez Linares Senior Research Fellow},
A. Caccianiga\inst{6},
R. Della Ceca\inst{6},\\
O. Gonz\'alez-Mart\'in\inst{11},
G. Lamer\inst{10},
P. Severgnini\inst{6}
}

   \offprints{P. Esquej \\ \email{pilar.esquej@cab.inta-csic.es}}

   \institute{
$^{1}$	Centro de Astrobiolog\'ia (INTA-CSIC), ESAC Campus, PO Box 78, 28691 Villanueva de la Ca\~{n}ada, Spain\\  
$^{2}$	Dept. of Physics and Astronomy, Leicester University, Leicester LE1 7RH, U.K. \\
$^{3}$	Instituto de F\'isica de Cantabria (CSIC-UC), Avenida de los Castros, 39005 Santander, Spain\\
$^{4}$  Departamento de F\'{\i}sica Moderna, Universidad de Cantabria, Avda. de Los Castros s/n, 39005 Santander, Spain\\
$^{5}$	Mullard Space Science Laboratory, University College London, Holmbury St. Mary, Dorking, Surrey RH5 6NT, UK\\
$^{6}$	INAF -- Osservatorio Astronomico di Brera, via Brera 28, 20121 Milan, Italy\\
$^{7}$	SRON - Netherlands Institute for Space Research, Sorbonnelaan 2, 3584 CA, Utrecht, The Netherlands\\
$^{8}$	European Southern Observatory, Karl-Schwarzschild-Stra$\beta$e 2, 85748 Garching bei M\"{u}nchen, Germany\\
$^{9}$	University of California, San Diego, Center for Astrophysics \& Space Sciences, 9500 Gilman Drive, CA 92093-0424, USA\\
$^{10}$	Leibniz-Institute for Astrophysics Potsdam (AIP), An der Sternwarte 16, 14482 Potsdam, Germany\\
$^{11}$	Instituto Astrof\'isico de Canarias,  (IAC), C/V\'{\i}a L\'{a}ctea, s/n, E-38205, La Laguna, Tenerife, Spain\\
}



  \abstract
   {}
   {This programme is aimed at obtaining one of the largest X-ray selected samples of identified active galactic nuclei to date in order to characterise such a population at intermediate fluxes, where most of the Universe's accretion power originates. We present the \xmm\ Wide Angle Survey (XWAS), a new catalogue of almost a thousand X-ray sources spectroscopically identified through optical observations. }
  {A sample of X-ray sources detected in 68 \xmm\ pointed observations was selected for optical multi-fibre spectroscopy. Optical counterparts and corresponding photometry of the X-ray sources were obtained from the SuperCOSMOS Sky Survey. Candidates for spectroscopy were initially selected with magnitudes down to $R\sim$21, with preference for X-ray sources having a flux $F_{0.5-4.5\,{\rm keV}}\geq$10$^{-14}$~\fluxunit. Optical spectroscopic observations performed at the Anglo Australian Telescope Two Degree Field were analysed, and the derived spectra were classified based on optical emission lines.}  
   {We have identified through optical spectroscopy 940 X-ray sources over $\Omega\sim$11.8 deg$^2$ of the sky. Source populations in our sample can be summarised as 65\% broad line active galactic nuclei (BLAGN), 16\% narrow emission line galaxies (NELGs), 6\% absorption line galaxies (ALGs) and 13\% stars. An active nucleus is likely to be present also in the large majority of the X-ray sources spectroscopically classified as NELGs or ALGs. Sources lie in high-galactic latitude ($\vert{\rm b}\vert > $ 20~deg) \xmm\ fields mainly in the southern hemisphere. Due to the large parameter space in redshift ($0\leq\,z\,\leq4.25$) and flux ($10^{-15}\leq\,F_{0.5-4.5\,{\rm keV}}\leq$\,10$^{-12}$~\fluxunit) covered by the XWAS this work provides an excellent resource to further study subsamples and particular cases. The overall properties of the extragalactic objects are presented in this paper. These include the redshift and luminosity distributions, optical and X-ray colours and X-ray-to-optical flux ratios.} 
   {}

   
   \authorrunning{P. Esquej et al.}
   \titlerunning{The \xmm\ Wide Angle Survey}
   \maketitle


\section{Introduction}\label{sec:sec1}
According to synthesis models, the growth of supermassive black holes (SMBHs) by accretion over cosmic time is recorded in the X-ray source population which produces the cosmic X-ray background (CXB). Hence, X-ray surveys can be used to constrain the epochs and environments in which SMBHs formed and evolved \citep{Alexander2003,Hasinger2005,Gilli2007}. X-ray surveys with high sensitivities and good spatial and spectral resolution are essential for studying the properties of the bulk of X-ray sources. Deep pencil beam surveys are able to detect sources down to very faint fluxes, therefore allowing the detection of typical sources in the sky and contributing to the picture of the early Universe. Among these we can find the \chandra\ deep field (CDF) surveys \citep[e.g.][]{Alexander2003,Tozzi2006,Luo2008,Xue2011}, and the \xmm\ deep surveys of the CDF--South (Ranalli et~al. in prep.) and the Lockman Hole \citep{Hasinger2001,Mainieri2002,Mateos2005}. All-sky surveys like the \emph{ROSAT} All-Sky Survey \citep{Voges1999} with shallow exposures but with large sky coverage can observe rare objects with small surface number density, and are able to unveil the bright end of the luminosity function. 

Serendipitous surveys, covering X-ray fluxes between 10$^{-12}$ and 10$^{-15}$~\fluxunit, fall in between especially designed all-sky programs and dedicated pointed observations. Sources at these intermediate fluxes are responsible for a large fraction of the X-ray background, as they sample the region around the break in the X-ray source counts \citep{Carrera2007,Mateos2008}. A number of campaigns have been dedicated to the optical-to-radio characterization of X-ray sources selected at different X-ray depths like the XB\"{o}otes survey \citep{Murray2005,Brand2006}, the Extended Groth strip Survey \citep[EGS][]{Georgakakis2006}, the HELLAS2XMM survey \citep{Cocchia2007,Fiore2003}, the \xmm\ and \chandra\ surveys in the COSMOS field \citep{Hasinger2007,Elvis2009,Brusa2010} and the Bright Ultra-Hard \xmm\ Survey \citep[BUXS;][]{Mateos2012}.

Data from \xmm\ observations have been used to create the largest X-ray catalogue ever produced, the 2XMMi-DR3 \citep{Watson2009}. Specific complete subsets of sources have already been used to investigate cosmological properties such as the X-ray log $N$\,--\,log\,$S$  distributions and the angular clustering of X-ray sources \citep{Mateos2008,Ebrero2009}. Optical imaging and spectroscopy of well-defined datasets from selected \xmm\ fields have been obtained in order to characterise their X-ray source populations. The \xmm\ Survey Science Centre (SSC) follow-up and identification (XID) programme is outlined in \citet{Watson2001}. Their goals include the detailed characterisation of the dominant X-ray source populations and the discovery of new classes of probable rare sources. The outcome of these samples is currently being used to establish statistical identification criteria in order to characterise the complete XID database \citep{Pineau2011}. The \xmm\ Bright Serendipitous Survey \citep[XBS;][]{DellaCeca2004,Caccianiga2008} and X-ray luminosity function \citep{DellaCeca2008,Ebrero2009b} have been published together with the XID Medium Survey catalogue \citep[XMS;][]{Barcons2007}, the X-ray source counts and the angular clustering \citep{Carrera2007}, and the X-ray spectral analysis of the active galactic nuclei \citep{Mateos2005}.

The \xmm\ Wide Angle Survey (XWAS) presented here is part of the follow-up programme conducted by the \xmm\,collaboration. It complements previous surveys in providing a qualitative picture of the X-ray sky. It yields optical and X-ray characterisation of $\sim$1000 objects mainly in the southern hemisphere. Spectroscopic optical observations have been used to provide the redshift and classification of the sources. Published papers based on sources drawn from the XWAS sample include detailed X-ray spectral analysis of the identified broad line active galactic nuclei  (BLAGN) \citep{Mateos2010}, stacking of all XID BLAGN spectra \citep{Corral2008} and a study and X-ray stacking of type II QSOs \citep{Krumpe2008}.

This paper is structured as follows: in Sect.~\ref{sec:sec2} we define the XWAS. In Sect.~\ref{sec:sec3} we discuss the multi-band optical imaging and spectroscopic observations conducted on the \xmm\ target fields. Sect.~\ref{sec:sec4} gives details on the source spectroscopic classification scheme, followed by the counterpart selection procedure in Sect.~\ref{sec:sec_CS}. Section~\ref{sec:sec5} presents the overall source populations. Section~\ref{sec:catalogue} describes the catalogue columns and how to obtain the data. Finally, Sect.~\ref{sec:sec6} summarises our main results.

Throughout this paper a $\Lambda$CDM cosmology of ($\Omega_{\rm m}$,~$\Omega_{\Lambda}$)~=~(0.3,~0.7) and ${H}_{0}$~=~70~ ${\rm km}~{\rm s}^{-1}~{\rm Mpc}^{-1}$ will be assumed. Optical magnitudes are given in the Vega system.

\section{The sample}\label{sec:sec2}
The \xmm\  satellite features high spectral resolution and excellent sensitivity due to the large collecting area of its mirrors coupled with the high quantum efficiency of the EPIC detectors \citep{Jansen2001}. It provides a significant detection of serendipitous sources in addition to the original target. The purpose of the XWAS project is to create a catalogue of X-ray sources detected by \xmm\ with optical identifications including redshift and classification based on our own optical multi-fibre spectroscopy.

The \xmm\ field selection criteria for the XWAS prioritised those observations with adjacent or overlapping coverage, to take optimum advantage of the field of view of the spectrograph used for the optical observations. The Galactic latitude was restricted so only those fields with $\vert{\rm b}\vert > $ 20~deg were selected in order to avoid high Galactic absorption and source confusion in the Galactic plane. We also required that all \xmm\ observations had total exposure times of $>\,10$\,ks and that they were performed in Full Frame mode in the EPIC cameras with thin or medium filters.

With these properties, X-ray objects were originally selected from 68 spatially distinct pointing observations, carried out by \xmm\ between June 2000 and May 2003. After the source selection, optical spectroscopic campaigns were performed. The Galactic absorbing column density along the line of sight for the selected observations is always $<2\times10^{21}\,{\rm cm}^{-2}$, which minimises non-uniformities introduced by large values of the Galactic $N_{\rm H}$. The median Galactic absorption is $3\times10^{20}~{\rm cm}^{-2}$. Of the total 5675 serendipitous X-ray sources found in the 68 exposures, $\sim$3000 objects were selected for optical spectroscopy (see Sect.~\ref{sec:sec3} for selection criteria). 

To derive the X-ray selection function, we have estimated the sky coverage as a function of the X-ray flux using empirical sensitivity maps for every observation \citep{Carrera2007,Mateos2008}. For each observation, we have estimated the minimum detectable count rate at each position of the field of view (FOV) for a threshold in detection significance (we have selected $\mathcal{L}\geq6$), taking into account the effective exposure and background level across the field of view. For a full description please see \citet{Carrera2007,Mateos2008}. For the count rate to flux conversion, which depends on the camera, observing mode, filter and spectral model we have used the energy conversion factors published in the 2XMMi-DR3 documentation\footnote{\href{http://xmmssc-www.star.le.ac.uk/Catalogue/2XMMi-DR3/UserGuide\_xmmcat.html\#ProblECFs}{\tt http://xmmssc-www.star.le.ac.uk/Catalogue/2XMMi-DR3}}.

\begin{figure}

\centering
\resizebox{1.0\hsize}{!}{\rotatebox[]{90}{\includegraphics{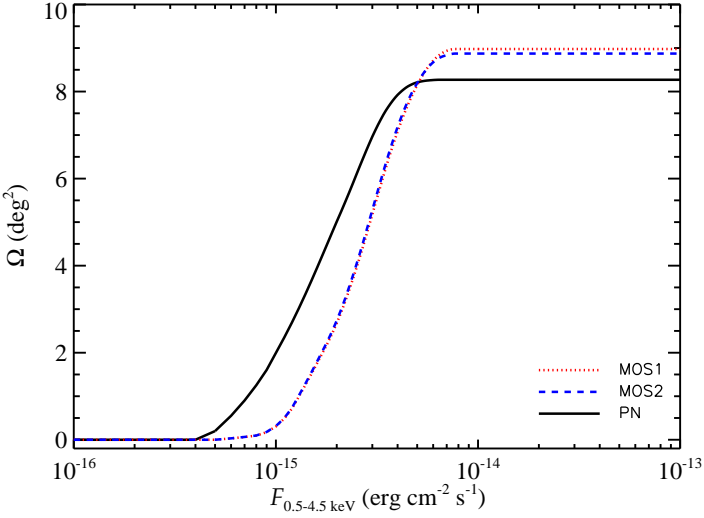}}}

\vspace{-1cm}
\caption[Sky coverage]{Distribution of the sky coverage as a function of X-ray flux in the 0.5--4.5\,keV energy band for the different cameras. The XWAS covers a total sky area of $\sim$11.3~deg$^2$, calculated using all EPIC cameras after correcting for overlaps.}
\label{fig:skyarea}
\end{figure}

The dependence of the sky coverage on the flux for the various cameras in the 0.5--4.5\,keV energy band is shown in Fig.~\ref{fig:skyarea}. The individual cameras survey $\sim$9\,deg$^2$ and $\sim$8\,deg$^2$ for the MOSs and pn respectively given that not all fields are observed by the three instruments. The fields targeted by the XWAS cover a net sky area of $\Omega\sim$11.3~deg$^2$, calculated using all three EPIC cameras after correcting for overlaps. In calculating the source counts we have only used flux levels at which sources are detectable over at least 1 deg$^{2}$ of sky. This constraint has been imposed to prevent uncertainties in the source count distributions due to low count statistics, and to avoid inaccuracy in the sky coverage calculation at the very faint detection limits.

\section{Source identification}\label{sec:sec3}

\subsection{Optical imaging}\label{sec:sec3.1}
Optical images are available for all our fields in the SuperCOSMOS Sky Survey \citep{Hambly2001}.
The SuperCOSMOS data primarily originate from scans of the UK Schmidt and Palomar POSS II blue, red and near-IR sky surveys. The ESO Schmidt $R$ and Palomar POSS-I $E$ surveys have also been scanned to provide an early epoch red measurement. The database hosting the SuperCOSMOS Science Archive (SSA) contains two main tables that hold the object catalogues. The \emph{Detection} table contains a list of detected objects on each scanned plate. Detections on the different plates are merged into a single source catalogue, the \emph{Source} table, that contains multi-colour data, given that each field is covered by four plates in passbands $B_{J}$, \emph{R} and \emph{I} (with \emph{R} being covered twice at different times, namely \emph{R1} and \emph{R2}). Extensive details on the surveys, the scanning process and the raw parameters extracted can be found under the SuperCOSMOS Sky Survey pages\footnote{\href{http://surveys.roe.ac.uk/ssa/index.html}{\tt http://surveys.roe.ac.uk/ssa/index.html}} and in \citet{Hambly2001}.

Our sample comprises both point-like and extended optical objects. The \emph{classMag} SuperCOSMOS parameters have been used as magnitudes in the different bands all converted to the Vega magnitude scale, hereafter and in the published tables of XWAS data. The most accurate photometric measurement from the plates depends on image morphology, and two different calibrations are applied to extended and point-like sources. However, one should note that the image classifier is not perfect and sometimes \emph{classMag} may not be the best photometric estimator for a given object. In the published XWAS catalogue we provide the SuperCOSMOS identifier so the user can return to the original tables and select either point-like or extended magnitudes if preferred. A 0.3\,mag uncertainty can be adopted for SuperCOSMOS magnitudes \citep{Hambly2001b}. These uncertainties account for the photometric accuracy of the plates and the different photographic emulsions used at different epochs. However, $B_{J}-$\emph{R} colours are expected to have an accuracy of $\sim$0.12 mag \citep[for details, see][]{Hambly2001}. Note that when using \emph{R} band photometry, \emph{R2} has been preferred due to its higher signal-to-noise ratio and better calibration when compared to \emph{R1} as noted in the documentation.

Detections in the SuperCOSMOS $B_{J}$ band have been primarily used for cross-correlation with the X-ray source positions due to their smaller position errors. Optical candidate counterparts were originally selected to be up to 10~arcsec from the X-ray position, although for the construction of the final catalogue a more stringent limit was imposed (see Sect.~\ref{sec:sec_CS}).  If no match was found, \emph{R2},  \emph{R1} or \emph{I} (in that hierarchical order) have been used for the counterpart selection. Further selection criteria imposed for the spectroscopic observations will be detailed in the following section.

\subsection{Optical spectroscopy}\label{sec:sec3.2}
We obtained optical multi-fibre spectroscopy of the X-ray sources with the Anglo Australian Telescope (AAT) Two Degree Field \citep[2dF;][]{Lewis2002}. Sources with X-ray counterparts having a 0.5$-$4.5\,keV flux $\geq$10$^{-14}$~\fluxunit\, were prioritised ($\sim$2500 sources). The selected energy range was chosen to maximise the \xmm\ EPIC sensitivity. This band is a good compromise between a broad passband (to favour throughput) and a narrow passband (to minimize non-uniformities in the selection function due to different source spectra). The 0.5~keV threshold was imposed to reject very soft photons and reduce the strong bias against absorbed sources occurring when selecting at softer energies. 

\begin{table}

\caption{Optical spectroscopic observations. RA and Dec refer to the field centre of the 2dF observations.}
\label{table:info}      
\centering                         
\begin{tabular}{|l|l|c|l|}
\hline
  \multicolumn{1}{|c|}{RA} &
  \multicolumn{1}{c|}{Dec} &
  \multicolumn{1}{c|}{Exposure} &
  \multicolumn{1}{c|}{Notes} \\
  & & \multicolumn{1}{c|}{(s)}&\\
\hline
  20:43:30.00 & --31:53:20.0 & 4800 & ADC problems\\
  01:40:40.00 & --67:51:00.0 & 4800 & poor focusing + aurora\\
  00:50:03.18 & --52:08:17.4 & 3600 & cloud\\
  21:38:00.00 & --43:05:14.0 & 1200 & dusty sky\\
  21:52:00.70 & --27:31:50.0 & 2400 & dusty sky\\
  22:52:30.00 & --17:45:00.0 & 4400 & \\
  22:16:00.00 & --17:15:00.0 & 3600 & \\
  02:25:07.97 & --05:02:26.8 & 4800 & \\
  02:23:52.10 & --03:49:00.6 & 3600 & poor seeing\\
  05:07:42.30 & --37:30:46.0 & 1800 & \\
  00:15:10.00 & --39:12:00.0 & 3440 & ADC problems, cloud\\
  21:04:25.00 & --11:51:00.0 & 4800 & \\
  23:02:50.00 & +08:45:00.0 & 3600 & ADC problems\\
  05:22:58.00 & --36:27:31.0 & 1800 & ADC problems, cloud\\
  21:51:55.60 & --30:27:53.7 & 3600 & \\
  01:34:00.00 & --40:36:31.0 & 3600 & cloud\\
  00:45:00.00 & --20:40:25.0 & 3600 & \\
  03:55:30.67 & +00:37:30.0 & 4200 & poor focusing + aurora\\
  03:36:36.00 & --25:33:25.0 & 1800 & \\
  01:52:53.00 & --13:50:00.0 & 4200 & ADC problems, cloud\\
  02:37:15.50 & --52:15:20.0 & 2250 & ADC problems, cloud\\
  23:16:14.00 & --42:32:50.0 & 3600 & \\
  03:37:38.00 & +00:28:40.0 & 3300 & \\
  05:05:20.00 & --28:49:05.0 & 3600 & \\
  04:09:11.23 & --71:17:41.9 & 3600 & \\
  03:39:03.50 & --35:26:30.0 & 3300 & \\
\hline\end{tabular}

\end{table}

Candidates for spectroscopy were initially selected above $R\sim$21, excluding the  targets of the \xmm\ observations. About 1200 objects ($\sim$21\% of all sources in the 68 \xmm\ observations) fulfilled these criteria. The 2dF provides more fibres per field than required for this programme. A significant fraction of the spectroscopic fibres were placed on lower probability counterparts, allowing for lower X-ray fluxes and fainter optical magnitudes (up to $R=21.66$) to be reached for a number of cases. Those X-ray sources with SuperCOSMOS counterpart offsets $>$~5~arcsec were entered with a low priority into the 2dF fibre positioning software, to allow for detection in case that they might be related to extended objects -- e.g. galaxy clusters.

We obtained optical spectroscopic observations for 27 2dF fields for the potential counterparts of a total of $\sim$3000 X-ray detections. General information for the observed fields is presented on Table~\ref{table:info}. Only one optical candidate per X-ray source could be observed  given that 2dF fibres cannot be positioned closer than 20\,arcsec from each other. Fibres, with a diameter of $\sim$2.1~arcsec, were placed at the positions of the optical counterparts derived from SuperCOSMOS. Observations of one hour per field were typically performed, normally split into in 3 exposures of 1200\,s to enable cosmic ray rejection. The XWAS 2dF spectroscopic observations provide an effective resolution of $\lambda/\delta\lambda\sim$600 over a wavelength range $\sim$3850--8250~\AA\ and reach a S/N of $\sim$5 at 5500~\AA\ for $V$=21\,mag. This is sufficient to provide a reliable object classification and redshift determination, together with a reasonable characterisation of the optical continuum shape for most of the objects. However, not all objects could be classified using this wavelength coverage and signal-to-noise ratio. Calibration lamp and flat-field exposures were taken before or after each science exposure, and observations of standards were performed in order to achieve the flux calibration of the targets. 
A number of exposures suffered from problems with the atmospheric dispersion corrector \citep[ADC][]{Lewis2002} or from non-optimal observing conditions such as cloud, poor seeing or the aurora. Spectra taken when the ADC was malfunctioning have distorted shapes due to wavelength-dependent light loss, particularly in the blue, but in many cases the spectra were still useful for identification.

The initial data reduction was carried out using the 2dF data reduction software\footnote{\href{http://www.aao.gov.au/2df/manual/UsersManual.pdf }{http://www.aao.gov.au/2df/manual/UsersManual.pdf}} (2dfdr). This included bias and dark subtraction, flat fielding, tram-line mapping to the fibre locations on the CCD, fibre extraction, arc identification, wavelength calibration, fibre throughput calibration and sky subtraction. Flux calibration, removal of the telluric absorption features, and improvement of the sky subtraction were performed with the IRAF\footnote{Image Reduction and Analysis Facility (IRAF) software is distributed by the National Optical Astronomy Observatories, which is operated by the Association of Universities for Research in Astronomy, Inc., under cooperative agreement with the National Science Foundation.} software package. Wavelength calibration accuracy is always better than 0.5\,\AA\ in the residuals. However, the flux calibration can only be considered as a calibration of the wavelength-dependence of the throughput, rather than as an absolute calibration, and even the relative flux calibration is not correct for spectra affected by the ADC problem. 

Henceforth, the fibre coordinates will be considered the reference position of our objects. 

\nocite{Lewis2002}

\section{Source classification}\label{sec:sec4}
Optical spectroscopy is crucial for determining the source type and redshift. Optical spectra have been screened and analysed in order to derive corresponding spectroscopic classification according to the following criteria. 

Extragalactic sources are classified as broad-line active galactic nuclei (BLAGN) when their optical spectra are characterised by the presence of at least one emission line with FWHM $> 1000$~km~s$^{-1}$, usually the H Balmer series, Mg~{\sc ii}, C~{\sc iii}], C~{\sc iv} and/or Ly\,$\alpha$. Those sources exhibiting emission lines which all have FWHM $<1000$~km~s$^{-1}$ are classified as narrow emission line galaxies (NELGs). We did not attempt any intermediate classification, therefore types 1 to 1.5 Seyferts are included within the BLAGN category. NELG comprise type 1.8 to 2, H~{\sc ii} galaxies, starburst galaxies, narrow  line Seyfert~1 galaxies and low ionisation nuclear emission-line regions (LINERs). Counterparts with pure absorption line spectra and a spectral shape corresponding to a galaxy are classified as absorption line galaxies (ALGs). Optical images were screened to look for possible evidence of a galaxy concentration typical of clusters, although our final sample does not include any of these. This is due to several factors: (1) the centroid of the X-ray detection did not fulfil the criteria for counterpart selection probably due to the extended nature of the objects, and (2) the software for X-ray source detection is optimised for point-like sources and misses very extended or low surface brightness objects.

We note that we cannot apply emission line diagrams for source characterisation, as typical emission lines used for that purpose (e.g. H$\alpha$) are usually shifted out from the observing window due to the restricted wavelength coverage of the spectroscopic observations. In addition, in some cases the host galaxy H$\beta$ absorption can mask any emission at that position and prevents us from using it as a useful AGN indicator.

Regarding the Galactic population, X-ray sources with a stellar optical spectrum are labelled as \emph{star}. A detailed study of the stellar population of this survey is beyond the scope of this paper. Most of them are expected to be active coronal stars showing X-ray spectra generally peaking at $\sim$1~keV and dominated by soft X-ray line emission, as found in the \xmm\ Galactic Plane Survey \citep{Motch2010}.

We obtained identifiable spectra for 1250 fibres. 2dF identified sources previously classified according to NED\footnote{The NASA/IPAC Extragalactic Database (NED) is operated by the Jet Propulsion Laboratory, California Institute of Technology, under contract with the National Aeronautics and Space Administration.} agree with our classification except for a few exceptions, some of them probably due to a different instrumental resolution or distinct criteria in the class determination (see Appendix~\ref{app:app1}).

\section{Counterpart selection}\label{sec:sec_CS}

Given that sources were originally selected from an early epoch processing of the X-ray data, we have performed the correlation of our reference fibre positions for the identified objects with the most recent version of the \xmm\ serendipitous source catalogue\footnote{\href{http://xmmssc-www.star.le.ac.uk/Catalogue/2XMMi-DR3/UserGuide\_xmmcat.html}{http://xmmssc-www.star.le.ac.uk/Catalogue/2XMMi-DR3}}, the 2XMMi-DR3. This has been done to take advantage of the significant improvements over the previous data processing system, so we can obtain a better parametrisation of the X-ray sources and the removal of possible spurious detections. Three \xmm\ fields were excluded from the 2XMMi-DR3 catalogue because they were seriously affected by high background flares, so the clean net exposure time was lower than the threshold used for the \xmm\ pipeline. These were independently processed by us following the same recipe as in the pipeline, and were included here for cross correlation with sources with identified optical 2dF spectra. Fibre positions were also re-cross matched against SuperCOSMOS to obtain the final source photometry\footnote{The last version of the SSA was released on June 2008.}.

Candidate counterparts derived from optical spectroscopy had to be either within 4 times the statistical error (at 90\% confidence) on the X-ray position determination or within 4 arcsec from the position of the X-ray source. This last criterion was used to accommodate any residual in the astrometric calibration of the X-ray EPIC images. This coincides with the overall astrometric accuracy found for the 2XMM catalogue \citep{Watson2009}, ensuring the X-ray/optical coincidence. A total of 963 sources fulfilled those restrictions.

After screening all spectroscopically identified sources, we concluded that for 14 cases the X-ray detection software detects a single source while several objects appear on the visual inspection of the X-ray image. These cases have been rejected so that the quality of the sample is not compromised. This is because the characterisation of such X-ray counterparts is ambiguous due to the contribution of emission from an additional object. In all cases the X-ray detection is extended and/or has low detection likelihood ($\lesssim$\,20, which is the threshold used for X-ray sources in e.g. \citet{Mateos2010}). Other sources have been removed after further screening due to the following reasons: the recorded X-ray emission is contaminated by source photons from the target of the observation (2 sources), the centre of the X-ray emission is coincident with a different optical source (3), the fibre is located between two optical sources (3), the optical source is located close to a very bright optical object (1).

\begin{figure}

\centering
\resizebox{1.0\hsize}{!}{\rotatebox[]{90}{\includegraphics{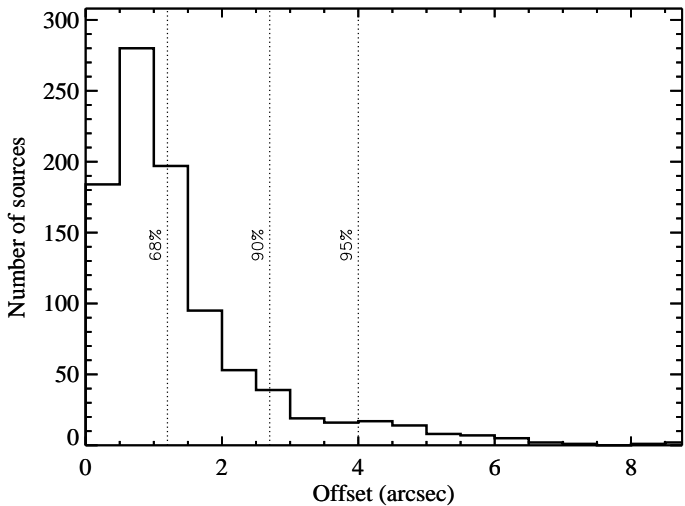}}}

\vspace{-1cm}
\caption[X-ray source distribution]{Histogram of the offsets between the optical source (fibre position) and the X-ray source centroid for spectroscopically identified objects. Vertical lines mark 68\%, 90\% and 95\% of the integration of the distribution. }
\label{fig:X-ray_hist}
\end{figure}

\begin{figure*}
\centering
\resizebox{\hsize}{!}{\rotatebox[]{90}{\includegraphics[clip=true]{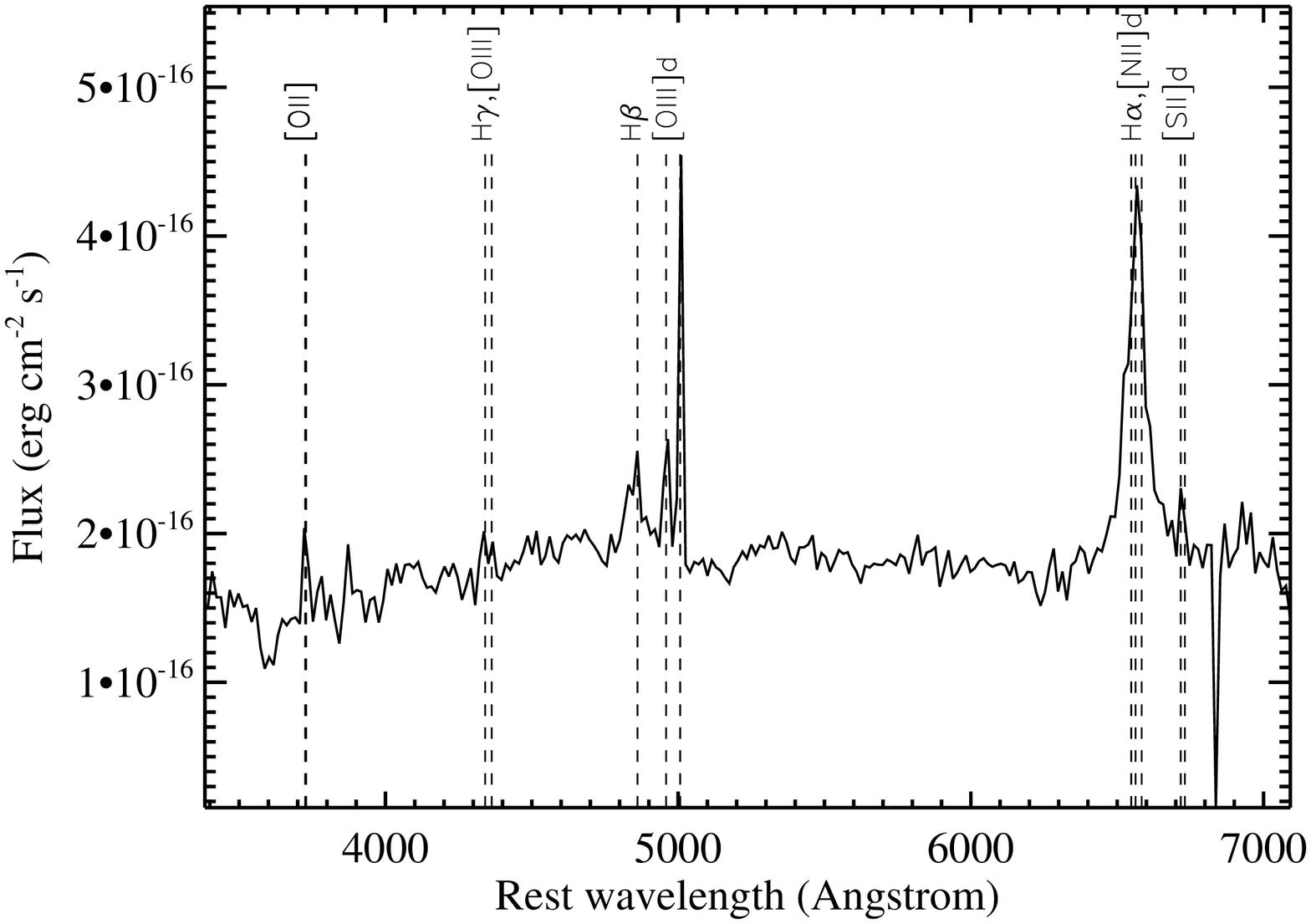}}
\rotatebox[]{90}{\includegraphics[clip=true]{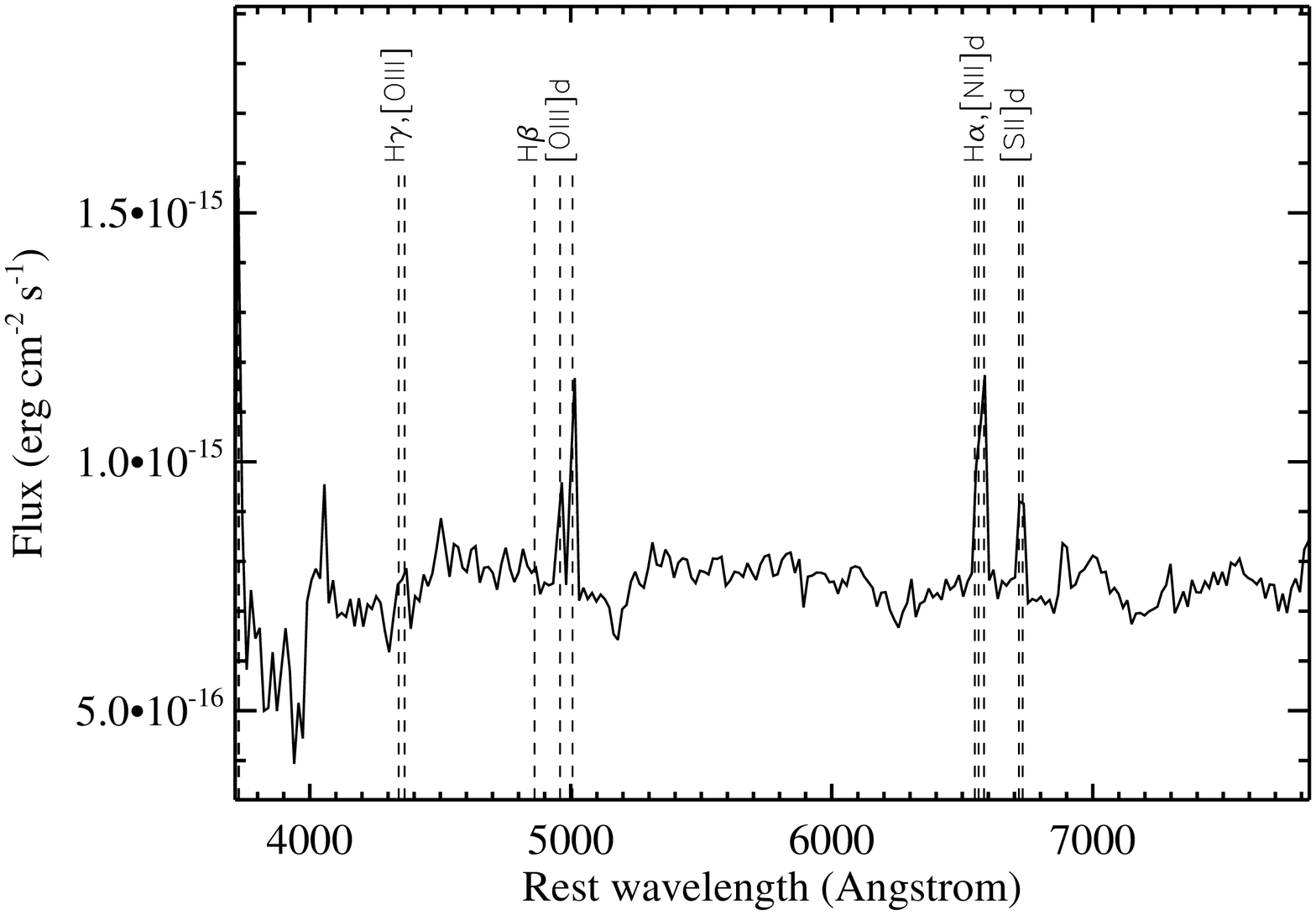}}}\vspace*{-1.5cm}
\resizebox{\hsize}{!}{\rotatebox[]{90}{\includegraphics[clip=true]{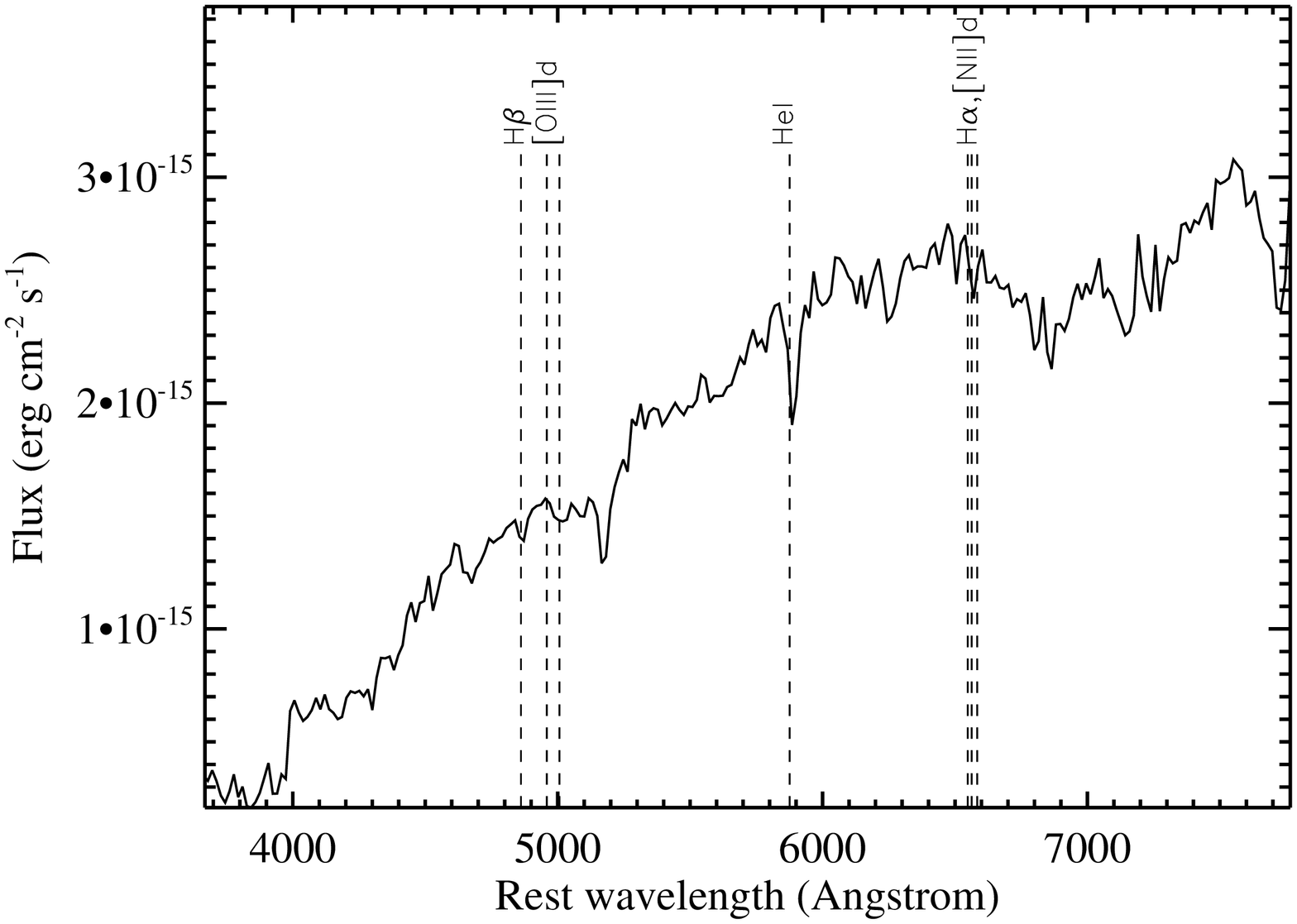}}
\rotatebox[]{90}{\includegraphics[clip=true]{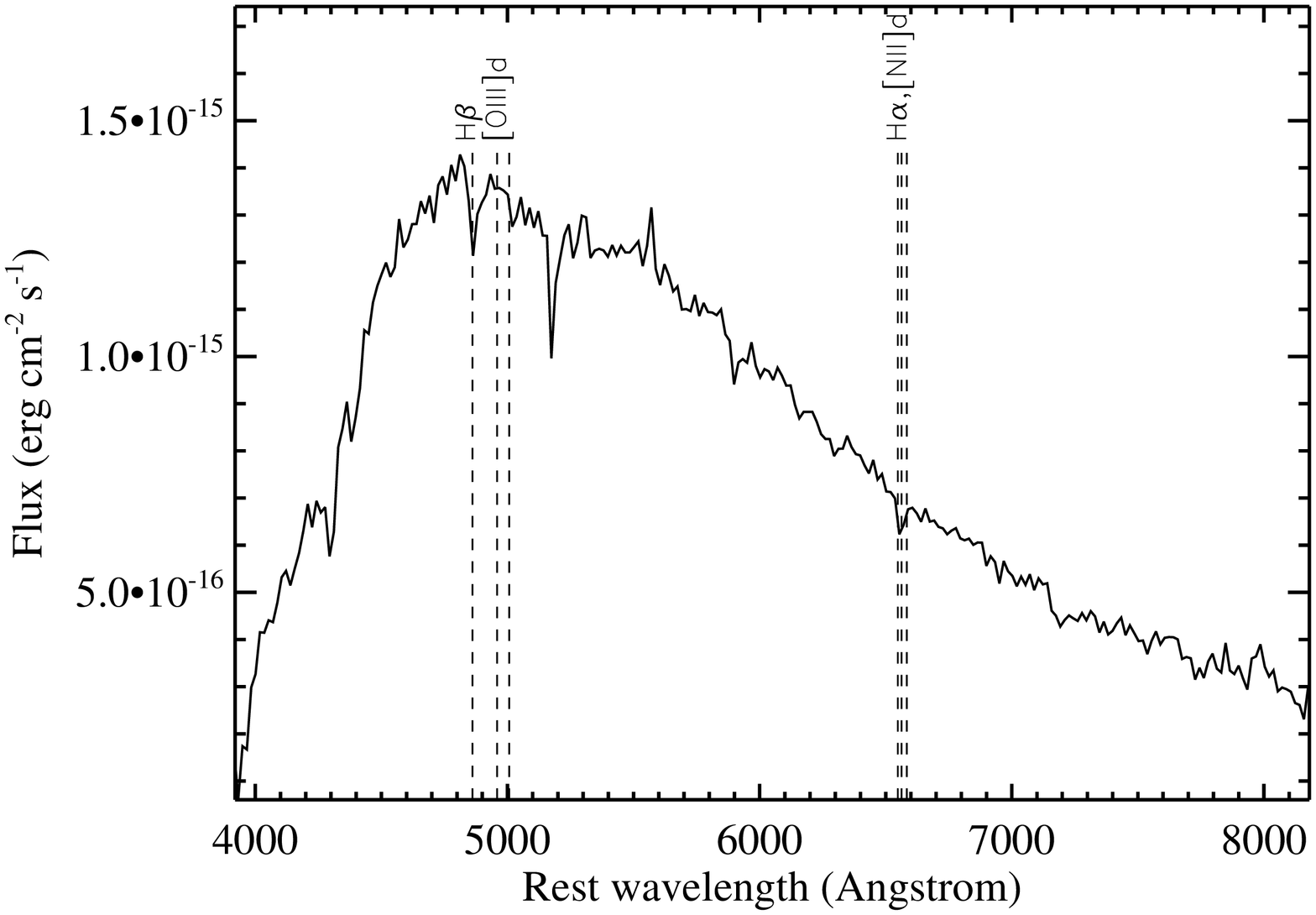}}}

\vspace*{-1.5cm}
\caption{Optical spectra of the different source types identified in the XWAS. Top left: BLAGN at z=0.152 -- XWAS J231658.6-423852. Top right: NELG at z=0.043 -- XWAS J033703.3-251456. Bottom left: ALG at z=0.056 -- XWAS J231756.3-421333. Bottom right: star -- XWAS J015319.4-135552. The positions of prominent emission and absorption lines are marked on the spectra.} 
\label{Fig:opt_spec}
\end{figure*}

The final XWAS catalogue includes 940 objects. We assume that the difference with the number of identified fibres (1250) is due to (a) our more restrictive assumptions in terms of optical-to-X-ray offset with respect to the original counterpart selection criteria, (b) differences in the \xmm\ software used for source determination as regards the version used for the original selection, and (c) the screening process. All these were needed in order to guarantee the highest possible quality catalogue. Fig.~\ref{fig:X-ray_hist} shows the histogram of the X-ray to optical angular offsets for spectroscopically identified sources. The integration of this distribution shows that for 68\%, 90\% and 95\% of the sample the optical counterpart lies closer than 1.2 arcsec, 2.7 arcsec and 4.0 arcsec respectively with respect to the X-ray position. Source populations in our sample can be summarised as 65\% BLAGN, 16\% NELGs, 6\% ALGs and 13\% stars. Fig.~\ref{Fig:opt_spec} shows examples of the different source types. We checked for spurious matches by cross-correlating almost 4000 random positions in the sky with SuperCosmos, the random positions obtained by shifting our source positions by $\pm$1 arcmin in RA and dec. We found contamination from spurious counterparts of only $\sim$5\% within 4 times the statistical error or 4\,arcsec.
 
\begin{figure}

\centering
\resizebox{1.0\hsize}{!}{\rotatebox[]{90}{\includegraphics{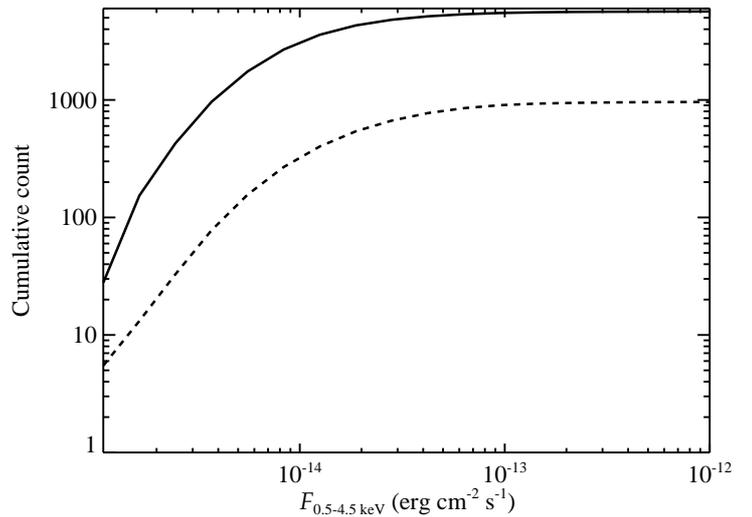}}}

\vspace{-1cm}
\caption[Selection function]{Cumulative count of X-ray sources as a function of flux in the 0.5--4.5\,keV energy band. The total number of detections in the original 68 \xmm\ observations (5675 sources) is represented by the solid line. Sources in the XWAS are shown by the dashed line.}
\label{fig:optselfunc}
\end{figure}

\begin{figure}

\centering
\resizebox{1.0\hsize}{!}{\rotatebox[]{90}{\includegraphics{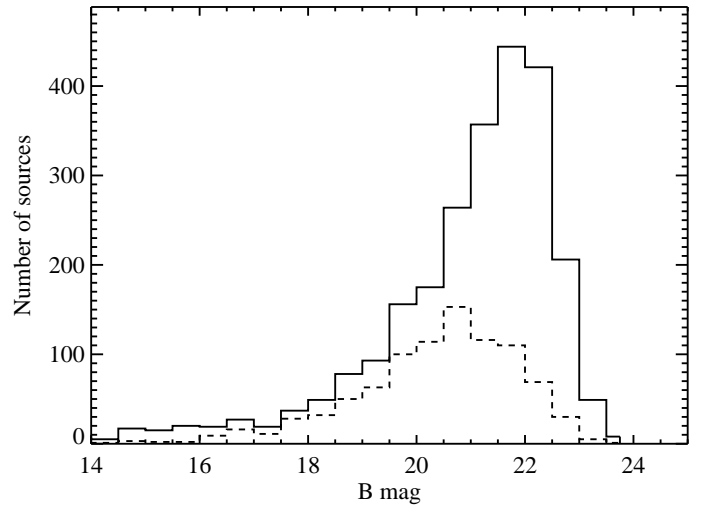}}}

\vspace{-1cm}
\caption{Distribution of the optical magnitude for all X-ray sources in the XWAS fields (solid line). The corresponding histogram of identified sources in the XWAS sample is shown by the dashed line).}
\label{fig:bmag}
\end{figure}

We have estimated the completeness of our sample by deriving the number of identified matches with respect to the total number of sources in the XMM-Newton fields (see Fig.~\ref{fig:optselfunc}). In addition, we show our spectroscopic success rate as a function of the optical magnitude. In Fig.~\ref{fig:bmag} we plot the distribution of $B_{J}$ magnitudes for the counterparts of all X-ray sources in the XWAS fields in contrast to the distribution for objects successfully identified. There we can see that at magnitudes brighter than $B_{J}$=20\,mag our spectroscopic identification rate is $\sim$80\,\%, while this is $\sim$30\,\% for 20$<B_{J}<$24.

We note that, given the SuperCOSMOS limiting optical magnitude, our sample could be biased towards bright objects. In order to check and quantify this limitation, we analysed additional observations in a few selected XWAS fields performed with the Wide Field Camera (WFC) on the Isaac Newton Telescope (INT). Typical exposure times of 600s were used for observations in the \emph{g$^\prime$} and \emph{r$^\prime$} Sloan Digital Sky Survey filters. This produced images with limiting magnitude for point-like sources down to \emph{r$^\prime$}$\sim$23$-$24 for $\sim$1$-$1.5~arcsec seeing, typical in our observing runs. Data reduction was performed following the WFC pipeline procedures under the Cambridge Astronomy Survey Unit\footnote{http://www.ast.cam.ac.uk/$\sim$wfcsur} (CASU). The WFC images were analysed using standard techniques including de-bias, non-linearity and flat field corrections \citep[see][for a full description]{GonzalezSolares2011}. Errors in magnitudes are assumed to be of 0.2\,mag.

Objects in the XWAS were cross-matched with detections in the WFC images. We used colour equations derived as in \citet{GonzalezSolares2011} to obtain red WFC magnitudes in the Vega system that were compared with the corresponding SuperCosmos counterparts. Magnitudes of both observatories agree quite well, with a mean difference of $\sim$0.1\,mag.  From the comparison, we expect up to 8\% of sources having fainter magnitudes due to our limiting optical magnitude, which is the fraction of sources with SuperCosmos counterparts but having an additional viable fainter match in the WFC.

\section{Overall characteristics of the source populations}\label{sec:sec5}
\begin{figure}

\centering
\resizebox{1\hsize}{!}{\rotatebox[]{90}{\includegraphics{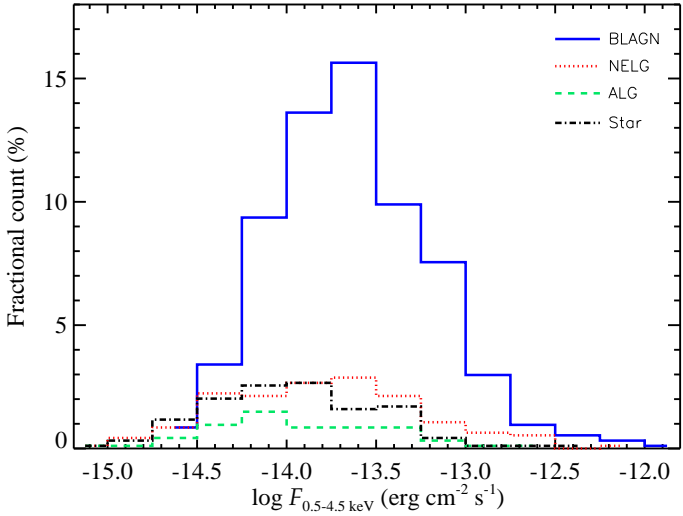}}}

\vspace{-1cm}
\caption{X-ray 0.5$-$4.5~keV flux distribution of sources with spectroscopic optical identification in the XWAS. Blue solid line: BLAGN; red dotted line: NELG; green dashed line: ALG; black dash-dotted line: star.}

\label{fig:Xray_flux_distrib}
\end{figure}

To illustrate the overall population sampled in the XWAS, Fig.~\ref{fig:Xray_flux_distrib} shows the flux distribution of the XWAS sources in the 0.5$-$4.5~keV band. We have used the EPIC fluxes appearing in the 2XMMi-DR3 catalogue. These are derived from the band count rates multiplied by a filter and camera-dependent energy factor \citep{Mateos2009}. This conversion assumes a spectral model consisting of a power-law with a continuum spectral slope $\Gamma$=1.7 and a photoelectric absorption $N_{\rm H} = 3 \times 10^{20}$ cm$^{-2}$ (for a general description, see the \xmm\ science survey centre memo, SSC-LUX TN-0059). Then, the EPIC flux in each band is the mean value of all cameras weighted by the errors. The model assumed in deriving the fluxes will be a fair representation for BLAGN, but less so for the other types of sources. We have included the correction for the Galactic column density using {\tt XSPEC} simulations of a power-law model ($\Gamma$=1.7) and the Galactic $N_{\rm H}$ of each individual source for the count rate to flux conversion. The estimated correction values are always less than a factor of 2. No attempt has been made to correct the fluxes for absorption of material intrinsic to the source.

\newpage

\subsection{Redshift and luminosity distributions}
\begin{figure}

\centering
\resizebox{1.0\hsize}{!}{\rotatebox[]{90}{\includegraphics{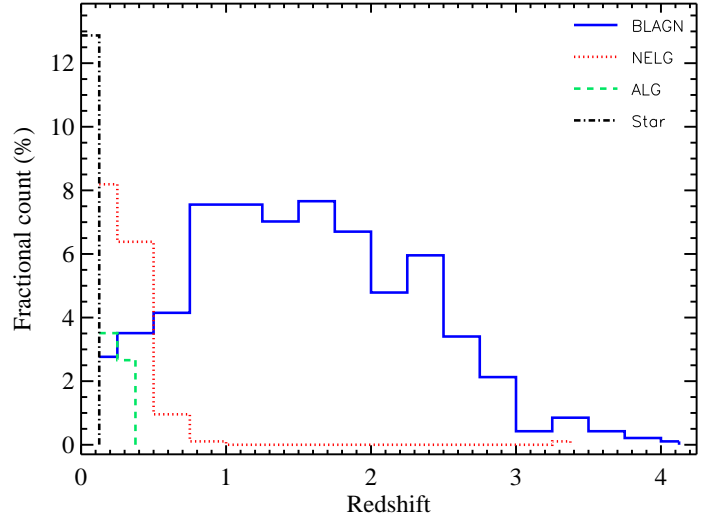}}}

\vspace{-1cm}
     \caption[Redshift distribution]{Redshift distribution of the XWAS objects coded by source type.}
        \label{fig:z_hist}
    \end{figure}

The redshifts were obtained as follows. First, each spectrum was manually inspected, classified, and an approximate redshift was determined from the wavelengths of the most prominent features, usually either emission or absorption lines. Then, the redshift was refined by cross-correlating the spectrum with a suitable template. The redshift distribution of the XWAS sources for different source types is displayed in Fig.~\ref{fig:z_hist}. From this histogram we can see that the distribution of BLAGN is broader than those of NELGs and ALGs. We derive the mean redshift of the BLAGN population to be $\avg{z}=1.5$, whereas the objects classified as NELGs and ALGs peak at lower redshift $\avg{z}=0.3$. Both values are comparable to those found in surveys with similar depths \citep[e.g.][]{Barcons2007}, while deeper surveys tend to find higher peak values for the non-BLAGN population \citep[e.g.][]{Silverman2005,Mateos2005,Xue2011}. We are able to detect BLAGN out to $z\sim 4$ in these medium depth observations. However, the large majority of those sources which are not BLAGN (all except for 4 NELGs) have $z<0.6$. The steep drop in the number of NELGs and ALGs above this redshift is almost certainly due to an optical selection bias, the combination of the optical faintness of these sources ($R\gtrsim$21) and the redshifting of the most easily-identifiable features outside the observing window.

\begin{figure*}

\centering

\resizebox{\hsize}{!}{\rotatebox[]{90}{\includegraphics{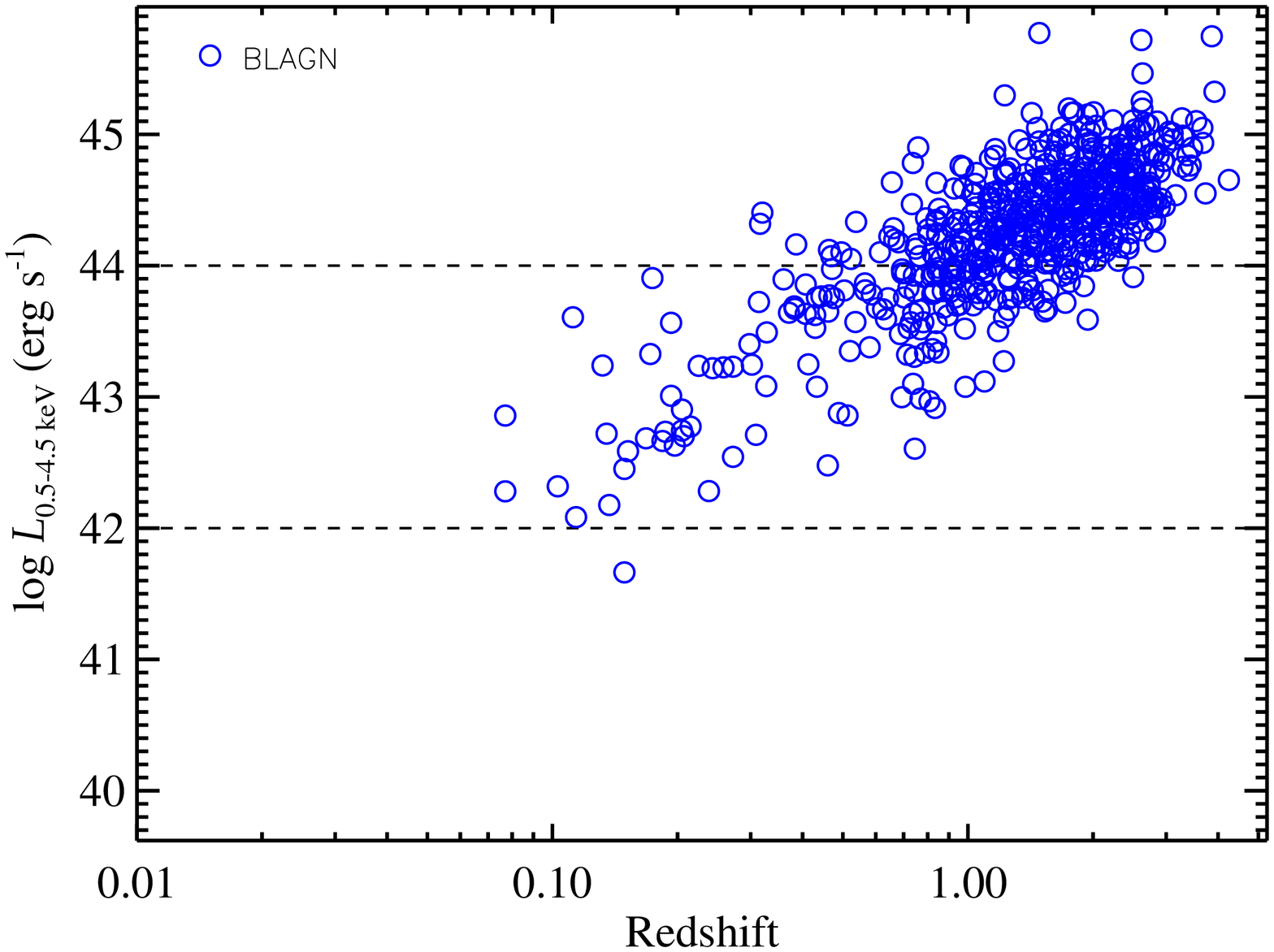}}
\hspace{0.5cm}   \rotatebox[]{90}{\includegraphics{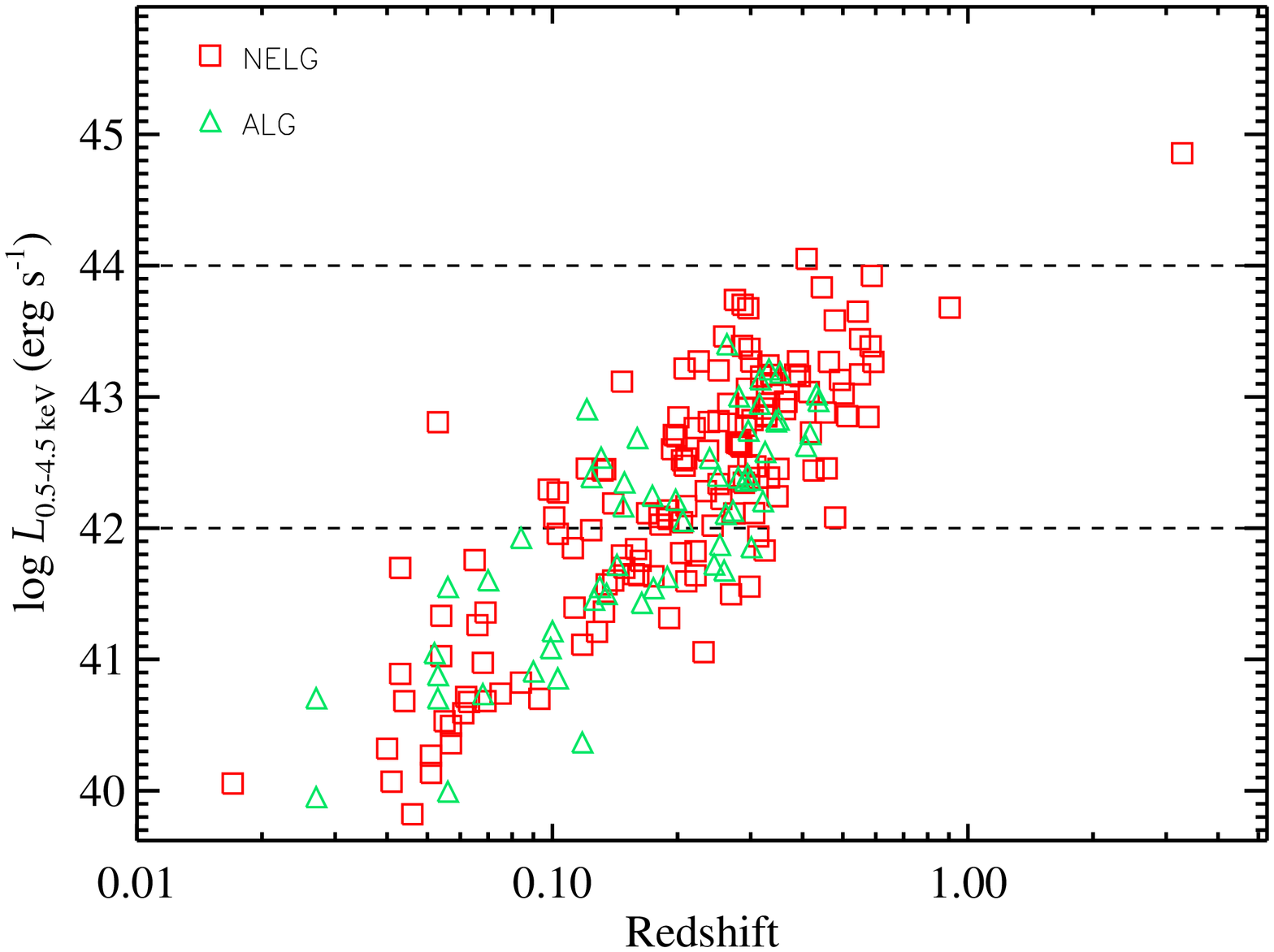}}}

\vspace{-1cm}
\caption{Observed X-ray luminosities in the 0.5$-$4.5~keV energy band as a function of redshift for extragalactic sources. Left panel: BLAGN. Right panel: NELG (red squares) and ALG (green triangles). }
\label{fig:Lum_vs_redshift}

\end{figure*}

X-ray luminosities in the 0.5--4.5\,keV energy band have been computed for the extragalactic objects of the sample using the redshifts obtained in our spectroscopic observations. In order to shift such values to a common rest-frame passband for all sources, a K-correction\ $k(z)$ \citep{Hogg2002} has been taken into account as

\begin{equation}
k(z)=(1+z)^{\Gamma-2}
\end{equation}

\noindent
where $\Gamma$=1.7 is the spectral photon index used for the count rate to flux conversion. X-ray luminosities (not corrected for intrinsic absorption) of the extragalactic sources as a function of redshift are presented in Fig.~\ref{fig:Lum_vs_redshift}. 
The sample contains both Seyfert-like AGN and Quasi Stellar Objects (QSO). This is because the overall luminosity distribution is centred around 10$^{44}$~\lumunit -- which is the quantity commonly used to separate Seyferts and QSOs -- where the bulk of the X-ray emission is produced as derived from the AGN X-ray luminosity function. Average properties of the extragalactic types are presented in Table~\ref{table:avg_prop}. Note that the large standard deviations in the table are indicative of the large parameter space covered by the XWAS.

  \begin{table}
\caption{General characteristics of the different spectroscopic types of extragalactic objects. Standard deviations are shown in brackets.}             
\label{table:avg_prop}      
\centering                         
\begin{tabular}{c c c c c}        

\hline\hline                 
\textbf{Class} & \textbf{$\avg{z}$} & \textbf{$\avg{F_{0.5-4.5\,\rm keV}}$}  & \textbf{$\avg{L_{0.5-4.5\,\rm keV}}$} \\
& & (\fluxunit) & (\lumunit) \\
\hline                        

   BLAGN & 1.5 (0.8) & 4 (8) $\times$10$^{-14}$ & 3 (5) $\times$10$^{44}$ \\     
   NELG  & 0.3 (0.3) & 4 (9) $\times$10$^{-14}$ & 1 (6) $\times$10$^{43}$  \\
   ALG   & 0.2 (0.1) & 2 (3) $\times$10$^{-14}$ & 3 (5) $\times$10$^{42}$  \\

\hline   
                                
\end{tabular}
\end{table}

Given that many of the traditional optical signatures of AGN (i.e. evident emission lines in optical spectra) are not present in obscured sources, high X-ray luminosity becomes our single discriminant for supermassive black hole accretion in a number of cases. Sources optically classified as NELGs with X-ray luminosities exceeding 10$^{42}$~\lumunit\ (73\,\% of the NELGs) are unlikely to be powered by star formation, and so this limit is placed in order to avoid objects dominated by star formation and X-ray binaries. Therefore they should be classified as type 2 AGN. In particular, the luminosity of the 3 NELGs exceed 10$^{44}$~\lumunit, and therefore qualify as type 2 QSOs by standard X-ray astronomy definitions. Two XWAS sources have been included in the sample of type 2 QSOs of \citet{Krumpe2008} solely based on their optical spectra.
It is worth mentioning that at the lower activity end $L_{\rm X}<10^{42}$~\lumunit, LINERs have been found to host active nuclei in a high number of cases \citep[80\%, e.g.][]{Gonzalez2009}. However, given that our classification lacks detail in that respect we cannot place further constraints on that particular class of activity.

There are 31 sources in the ALG class (52\,\%) with luminosities beyond 10$^{42}$~\lumunit.  Sources with such properties are commonly identified as X-ray bright optically normal galaxies \citep[XBONGs][]{Fiore2000,Barger2001,Comastri2002,Georgantopoulos2005}. They are usually found to host either heavily obscured or low luminosity AGN.  The lack of emission lines in the optical spectra is commonly attributed to several factors, such as the faintness of the AGN with respect to the host galaxy or a non appropriate wavelength coverage of the optical spectrum \citep[e.g.][]{Moran2002,Severgnini2003,Caccianiga2007,Krumpe2007}. Another argument that points to the presence of an active nucleus in NELGs and ALGs with luminosities higher than 10$^{42}$~\lumunit\ is that they have X-ray-to-optical flux ratios typical of AGN (see Sect.~\ref{sec:XO_ratio}).

\subsection{Optical colour distributions}
\begin{figure*}

\centering
\resizebox{\hsize}{!}{\rotatebox[]{90}{\includegraphics{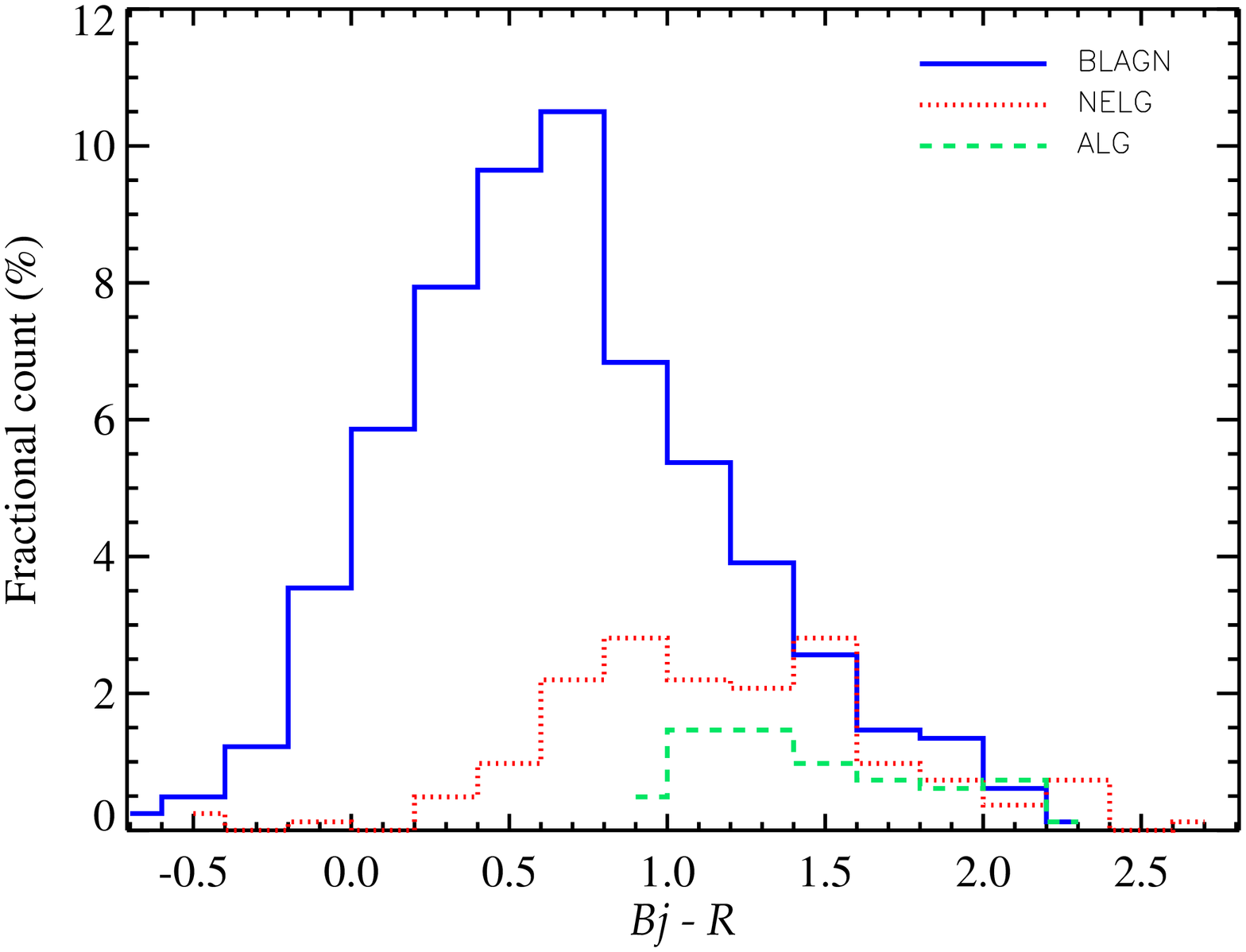}}
\hspace{0.5cm}   \rotatebox[]{90}{\includegraphics{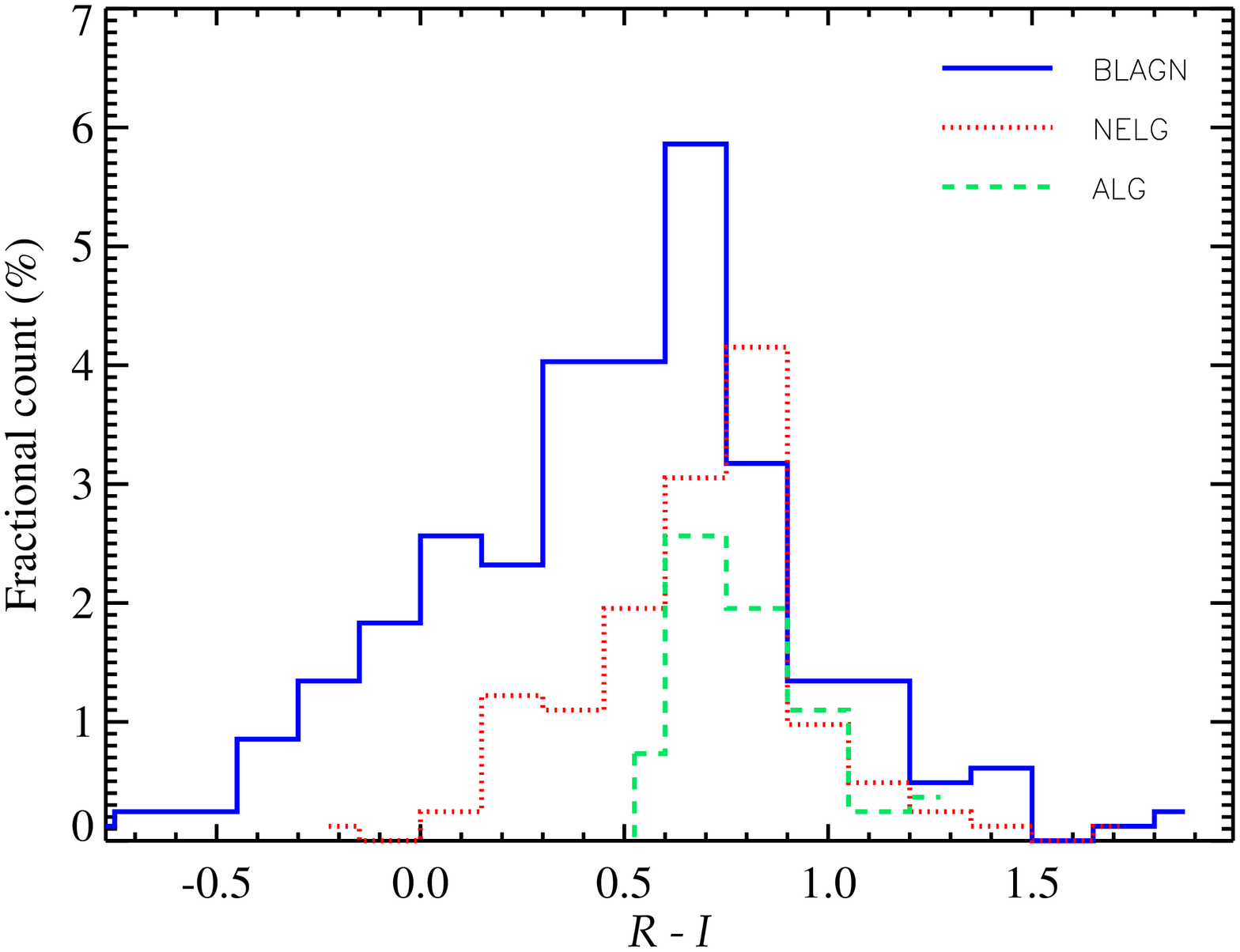}}}

\vspace{-1cm}
\caption{Histograms of the $B_{j}-R$ and $R-I$ colour distributions in the left and right panels respectively.}
\label{fig:hist_colour}

\end{figure*}

BLAGN are normally characterised by bluer optical colours than NELGs. This can also be seen in the average colours of our distributions, presented in Table~\ref{table:avg_colour}, where the average $\avg{{B_{j}-R}}$ is 0.96 for the former and 1.44 for the latter. In Fig.~\ref{fig:hist_colour} (left panel) we have plotted the $B_{j}-R$ colour distribution for the different extragalactic source types. For the $R-I$ colour histogram, shown in the right panel of Fig.~\ref{fig:hist_colour}, the average value for all populations is very similar, while a broader scatter on the distribution is seen for BLAGN in contrast to  NELGs and ALGs.
The Kolmogorov-Smirnov two-sample statistic has been estimated for the different colours. The small values of the significance level of the K-S test for the distinct populations (10$^{-11}$ and 0.04 for the $B_{j}-R$ and $R-I$ respectively) imply that the cumulative distribution of the two samples are significantly different.

\begin{table}
  
\caption{Mean optical colours  and hardness ratios with their corresponding standard deviations (in brackets) for the different spectroscopic types of sources.}             
\label{table:avg_colour}      
\centering                         
\begin{tabular}{c c c c c c c}        

\hline\hline                 
\textbf{Class} & \textbf{$\avg{\emph{B-R}}$} & \textbf{$\avg{\emph{R-I}}$} &
\textbf{$\avg{\rm HR}$}  \\

\hline                        

   BLAGN & 0.67 (0.53) & 0.48 (0.44) & --0.47 (0.26)\\     
   NELG & 1.17 (0.54) & 0.68 (0.28) & --0.23 (0.47)\\
   ALG & 1.46 (0.38) & 0.80 (0.18) & --0.62 (0.41)\\

\hline   
                                
\end{tabular}
\end{table}

\subsection{X-ray colour distribution}
X-ray spectral analysis of sources in our sample can only be performed in a limited number of cases. A crude spectral determination is available through the source X-ray colour, known as hardness ratio (HR). This is obtained by combining corrected count rates from different energy bands. The HR used here is defined as HR = $(S_{\rm h} - S_{\rm s})/(S_{\rm h} + S_{\rm s})$ where $S_{\rm h}$ and $S_{\rm s}$ are the count rates in the hard (2$-$10\,keV) and soft (0.5$-$2\,keV) bands respectively for a given source. By definition, $-1\leq HR \leq +1$. Values close to $-$1 indicate that the source has an extremely soft spectrum, while very hard or heavily absorbed sources are characterised by values close to +1.  

\begin{figure}

\centering
\resizebox{1\hsize}{!}{\rotatebox[]{90}{\includegraphics{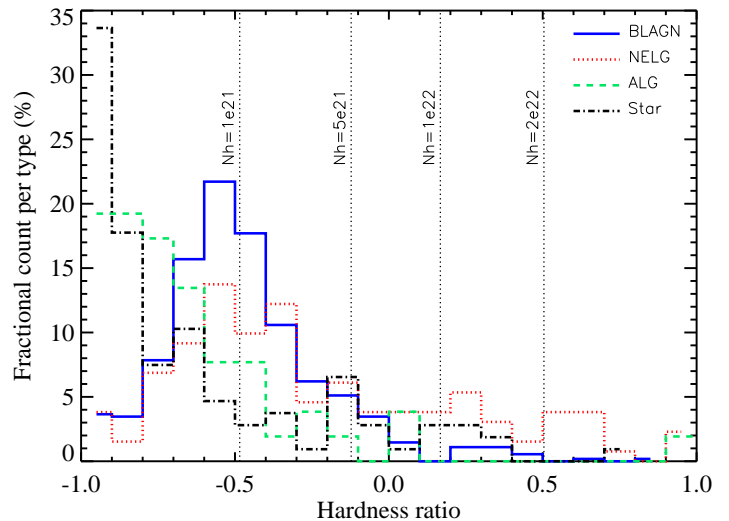}}}

\vspace{-1cm}
\caption{Hardness ratio distribution of sources in the XWAS. Vertical lines show the hardness ratios corresponding to a source of $z$=0.3 with a spectral slope of $\Gamma$=1.7 subject to different absorption values. }

        \label{fig:hr_hist}
\end{figure}

Fig.~\ref{fig:hr_hist} shows the EPIC-pn hardness ratio distribution (90\% of the total sample, i.e. sources observed with the EPIC-pn camera with detections in the individual soft and hard X-ray energy bands), where each population has been independently normalised. NELGs are expected to be absorbed sources, therefore we have simulated powerlaw spectra with X-ray slope of 1.7 and a variety of absorption values at the typical redshift of our sources  $\avg{z}$=0.3. The hardness ratios corresponding to those spectra are shown as vertical lines in the figure. On average, the softest sources are the stars, followed by ALGs, BLAGN and NELGs (see also Table~\ref{table:avg_colour}). 

ALGs and NELGs have very similar redshift distribution, so one can directly compare the luminosity distributions of the two populations. On average, we find that ALGs are less luminous than NELGs for the same redshift range (3$\times10^{42}$ vs 9$\times10^{42}$\,\lumunit). This, in addition to the fact that ALGs are less absorbed than NELGs, is an indication that the non-active optical appearance of the ALGs is most likely due to a host galaxy effect, i.e. the emission lines and AGN continnum  are outshone by the stellar continuum as also found in \citet{Moran2002,Severgnini2003,Mateos2005,Page2006}.

\subsection{X-ray-to-optical flux ratio}\label{sec:XO_ratio}
A classical approach extensively used in X-ray surveys as a proxy for detecting obscured sources is the \emph{so-called} X-ray-to-optical flux ratio ($f_{\rm X}/f_{\rm opt}\equiv{X/O}$) diagnostic diagram \citep{Maccacaro1988}. Previous analyses have shown that X-ray selected unobscured AGN have typical $X/O$ between 0.1 and 10 \citep[][and references therein]{Fiore2003}. Flux ratios below 0.1 are typical of stars and normal galaxies; and ratios higher than 10 would correspond to heavily obscured AGN (but not Compton-thick), high redshift galaxy clusters and extreme BL Lac objects.

Here, the X-ray flux is defined as the 0.5$-$4.5~keV flux not corrected for Galactic absorption (the correction is not significant for our sources). For the optical flux we have used that in the red band, computed as

\begin{equation}
\log(f_{\rm opt})=-0.4R+\log(f_{R_0}\delta\lambda)
\end{equation}

\noindent
where $f_{R_0}=1.74\times 10^{-9}$\ \fluxunit\  as the zero-point for \emph{R} \citep{Zombeck1990} and $\delta\lambda= 2200\, {\rm \AA}$ as the FWHM of the red filter. We prefer to use $R$\,=\,\emph{R2} for SuperCOSMOS sources (or $R1$ if there is not $R2$ magnitude available). Therefore, we find

\begin{equation}
\log(X/O)=\log(f_{\rm X})+0.4R+5.42
\end{equation}

\begin{figure*}

\centering

\resizebox{\hsize}{!}{\rotatebox[]{90}{\includegraphics{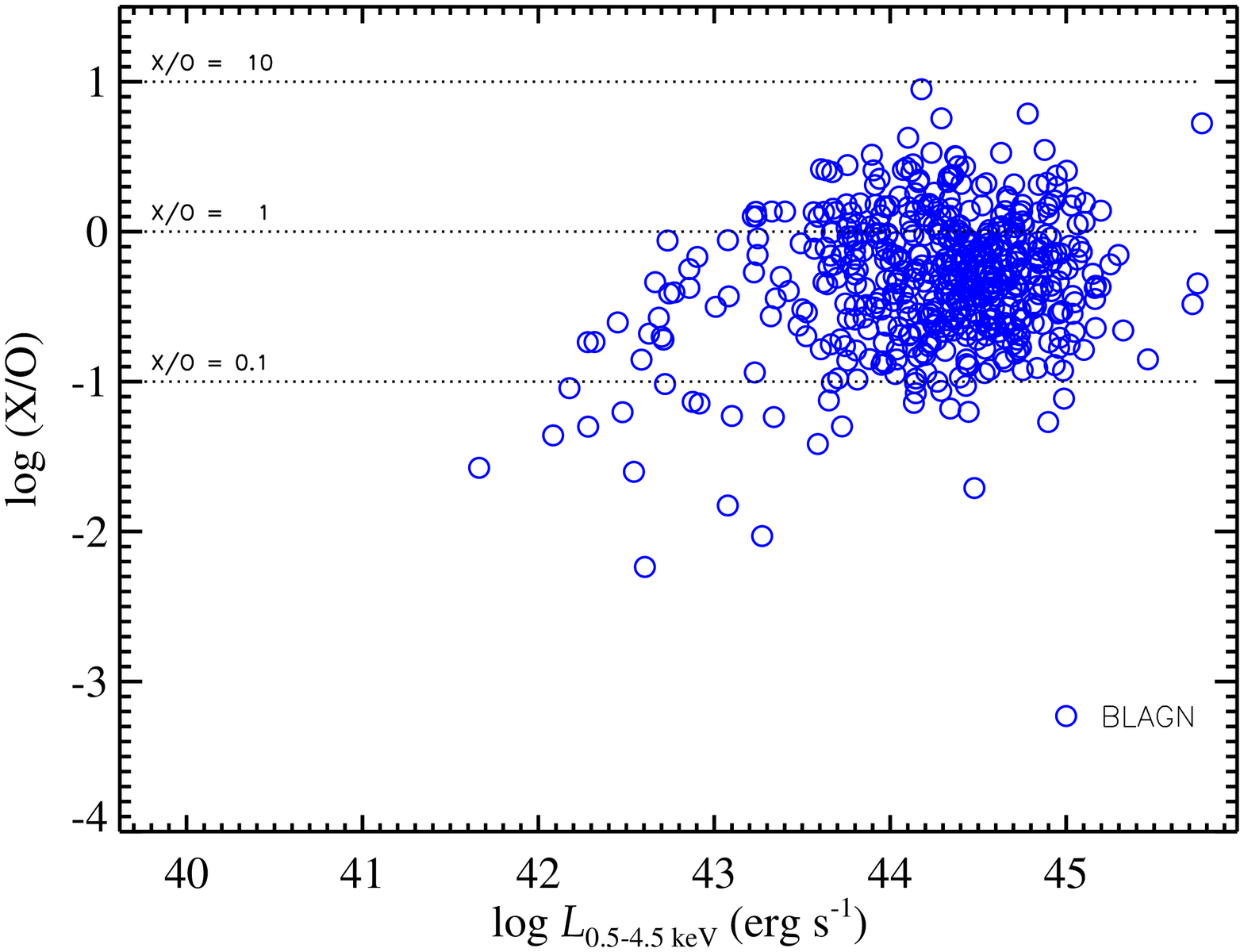}}
\hspace{0.5cm}   \rotatebox[]{90}{\includegraphics{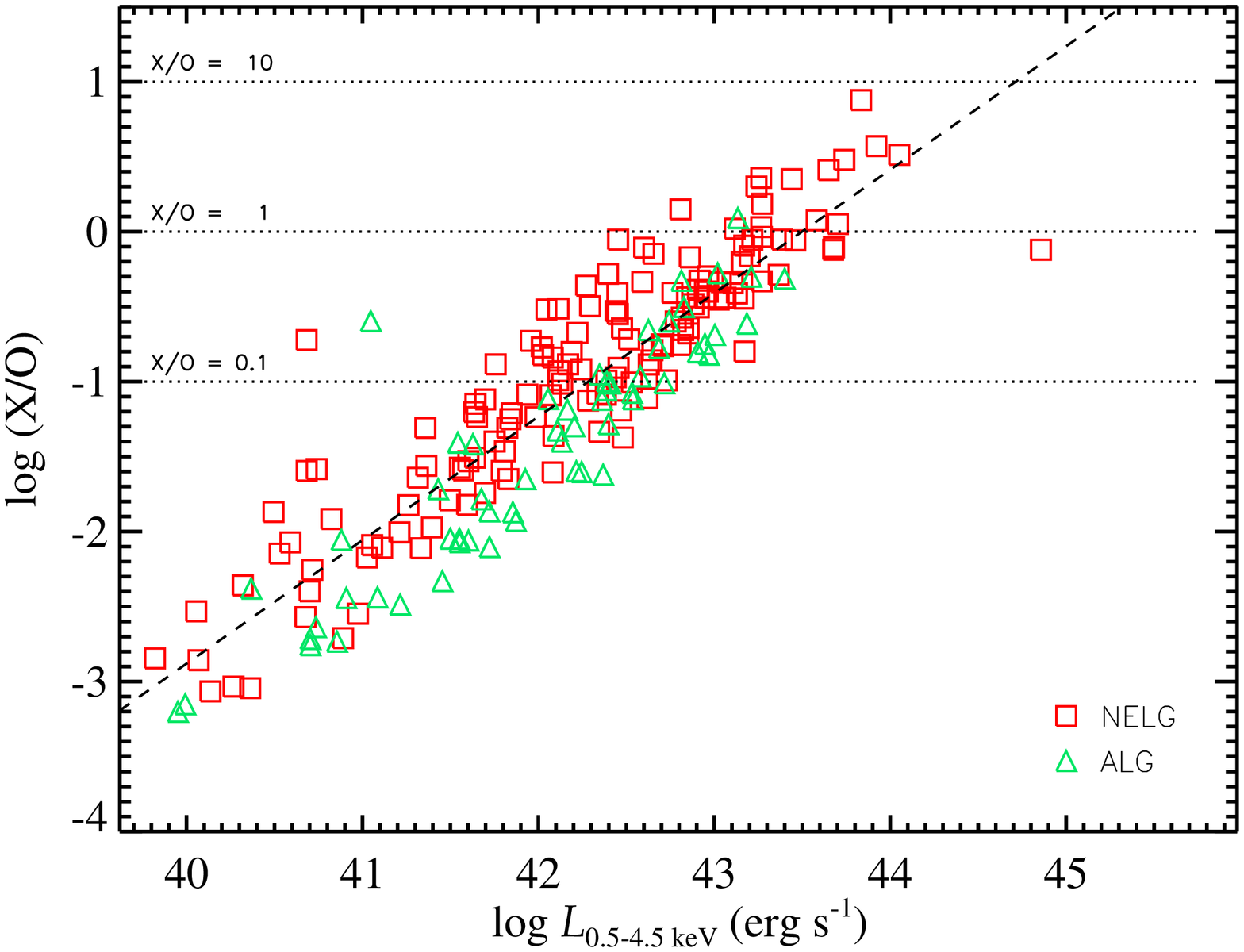}}}

\vspace{-1cm}
\caption{X-ray-to-optical flux ratio (X/O) as a function of the 0.5$-$4.5~keV X-ray luminosity for extragalactic sources. The horizontal dotted lines mark the level of $X/O$=10, 1 and 0.1 from top to bottom. Left panel: BLAGN. Right panel: NELGs and ALGs. The dashed diagonal line is the best linear regression between log\,($X/O$) and log($L_{0.5-4.5~{\rm keV}}$) for NELGs and ALGs with $L_{0.5-4.5~{\rm keV}}>10^{42}$~\lumunit (see text for details).}.
        \label{fig:FxFopt_Lum_v2}

\end{figure*}

Fig.~\ref{fig:FxFopt_Lum_v2} shows the X-ray-to-optical flux ratio as a function of the 0.5$-$4.5~keV X-ray luminosity for different extragalactic source types. The majority of sources detected in the 0.5$-$4.5~keV band have X-ray-to-optical flux ratios of typical AGN. We note that only one source in our sample has $X/O>10$. A small number was \emph{a priori} expected given that the initial threshold imposed on the optical flux of our sources was relatively high.  The BLAGN distribution does not show any trend. Due to the absence of broad emission lines in NELGs and ALGs, we expect their optical \emph{R} band emission to be dominated by the host galaxy given that the nuclear optical/UV emission is completely blocked (or strongly reduced). Therefore, $X/O$ is roughly a flux ratio between the nuclear X-ray and the host galaxy light emissions. As can be seen in the plot, there is a correlation between $X/O$ and the hard X-ray luminosity for non-BLAGN in such a way that higher luminosity sources tend to have higher $X/O$. The dashed diagonal line in Fig.~\ref{fig:FxFopt_Lum_v2} indicates the best linear regression only using detections 
between log\,($X/O$) and log($L_{0.5-4.5~{\rm keV}}$) for non-BLAGN with $L_{0.5-4.5~{\rm keV}}>10^{42}$~\lumunit (those expected to harbour a hidden AGN) and extrapolated to lower luminosities \citep[similar to that found in][]{Fiore2003}. 

\section{The catalogue}\label{sec:catalogue}

The catalogue consists of 940 entries, one per object. It contains information about the X-ray detection, optical imaging and optical spectroscopy for every object. Only a number of representative parameters of the 2XMM-DR3i and SuperCOSMOS archive appear in the XWAS. For additional information, we invite the user to search in the original tables. This can be done by looking for the IAUNAME and OBJID columns in the \xmm\ or SuperCOSMOS archives, respectively. 

The XWASNAME column represents the name assigned to the XWAS sources. They start with the prefix, XWAS, and then encode
the J2000 sky position of the X-ray object. Note that this coincides with the IAUNAME column in the 2XMM-DR3i aside from the prefix, except for the extra 50 sources not included in the \xmm\ catalogue due to the high background flares (Sect.~\ref{sec:sec_CS}). For those extra objects, we release a separate table with data from the X-ray pipeline processing similar to that in the \xmm\ catalogue. Some basic X-ray parameters directly extracted from the \xmm\ survey are included in the released XWAS catalogue. These are the source name, coordinates, positional error and flux in the total 0.5--4.5\,keV band. The X-ray luminosity has also been calculated using the redshift of the optical observations and included in the table.

For the optical information, we have an identifier derived from the 2dF observations, named OPTID. Fibre positioning and separation with respect to the X-ray position are also included, along with the object class and redshift derived from our analysis of the observations. From SuperCOSMOS we have included the OBJID, so the user can get all data from the original tables, plus the SuperCOSMOS magnitudes in the different bands when possible. When existing, we use the \emph{R2} magnitude in the \emph{R mag} column, otherwise, sources are flagged and we quote the \emph{R1} magnitude instead. A subset of columns of the XWAS catalogue is presented in Table~\ref{table:catalogue} and the complete catalogue will only be available in electronic form.

\section{Conclusions}\label{sec:sec6}
In this paper we have presented the strategy, production and overall characteristics of the new \xmm\ Wide Angle Survey. With almost a thousand sources selected in the 0.5--4.5\ keV energy band, this is one of the largest X-ray selected samples of spectroscopically identified AGN to date. The catalogue has a large scientific potential given the quality and high number of sources. It complements previous X-ray surveys to yield a qualitative picture of the X-ray sky.

The XWAS catalogue comprises 940 X-ray objects identified through optical observations performed by the 2dF multi-fibre spectrometer. Sources are distributed over $\Omega\sim$11.8 deg$^2$ in high-galactic latitude \xmm\ fields (--85~deg$< {\rm b} < $ --30\,deg). The large volume covered allows us to sample the bright end of the X-ray luminosity function. Source populations in our survey are 65\% BLAGN, 16\% NELGs, 6\% ALGs and 13\% stars. 

A high number of NELGs and ALGs are also presupposed to have an active nucleus given the X-ray luminosity and X-ray-to-optical flux ratios.
NELGs are the most absorbed sources in the survey as shown by their X-ray colours. Extragalactic sources with luminosities lower than 10$^{42}$\,\lumunit\ could also have high absorption and host AGN. Indeed, some works suggest that they can include Compton-thick AGN.

The sample presented here spreads over a large parameter space, in a region of the redshift-luminosity diagram poorly covered so far. The BLAGN sample extends out to redshift $4$, with an average of $\avg{z}$=1.5. The average value for NELGs is $\avg{z}$=0.3, and $\avg{z}$=0.2 for ALGs. This is in agreement with previous surveys with similar depth. As expected, the BLAGN appear bluer than those galaxies with narrow or no spectral emission lines.

A similar survey in terms of sky coverage and X-ray flux limits is the  XB\"{o}otes survey, with optical spectroscopy from the AGN and Galaxy Evolution Survey \citep[AGES;][]{Hickox2009}. The authors explore its multiwavelengh properties, but the radio, X-ray, and IR AGN samples only show a mild overlap. However, although it covers a similar X-ray luminosity range to the XWAS, the redshift sampling is quite limited $0.25\,<\,z\,<\,0.8$.

Due to the large covered volume, one can also perform stacking analyses of the X-ray data to determine the mean X-ray properties of different populations. In that context, \citet{Mateos2010} derived he mean properties of BLAGN, and the corresponding characteristics of the Fe K$_\alpha$ line were presented in \citet{Corral2008}.

The catalogue table can be accessed by direct download or via searches in the major astronomical databases. The XID results database contains additional information including direct links to X-ray and optical thumbnails and optical spectra\footnote{\href{http://xcatdb.u-strasbg.fr/xidresult/home}{http://xcatdb.u-strasbg.fr/xidresult/home}} that have been created for the present catalogue.

The results presented here can be an anticipation of what will be seen in future planned X-ray surveys. As an example, the XMM-XXL survey \citep{Pierre2011} will cover two extragalactic regions of 25\,deg$^2$ (at a depth of $5\times 10^{-15}$\,\fluxunit), and eROSITA \citep{Predehl2010} will perform an all-sky survey at a limiting flux of 10$^{-14}$\,\fluxunit).

\begin{table*}
\caption{Subset of the XWAS including the first 20 sources of the catalogue, ordered by right ascension. Null values have been substituted by 99.99.}             
\label{table:catalogue}      

\begin{tabular}{|l|r|r|l|c|r|l|l|l|l|}
\hline
  \multicolumn{1}{|c|}{XWASNAME} &
  \multicolumn{1}{c|}{$F_{\rm X}^{(1)}$} &
  \multicolumn{1}{c|}{log\,$L_{\rm X}^{(2)}$} &
  \multicolumn{1}{c|}{OptID} &
  \multicolumn{1}{c|}{Type} &
  \multicolumn{1}{c|}{$z$} &
  \multicolumn{1}{c|}{ScosID} &
  \multicolumn{1}{c|}{$B_{J}$ mag$^{(3)}$} &
  \multicolumn{1}{c|}{$R$ mag$^{(3)}$} &
  \multicolumn{1}{c|}{$I$ mag$^{(3)}$} \\

\hline
  XWAS J000002.7-251136 & 3.1 & 44.50 & X21207\_032 & BLAGN & 1.314 & 283502201403207 & 21.53 & 21.33 &99.99\\
  XWAS J000021.2-250812 & 2.2 & 44.05 & X21207\_051 & BLAGN & 0.995 & 283502201406064 & 20.41 & 20.89 &99.99\\
  XWAS J000022.8-251222 & 2.5 & 44.61 & X21207\_028 & BLAGN & 1.600 & 283502201402588 & 20.69 & 19.92 & 19.45\\
  XWAS J000025.3-251141 & 0.4 & 40.90 & X21207\_031 &   ALG & 0.090 & 283502201403144 & 17.18 & 16.35 & 15.98\\
  XWAS J000025.4-245421 & 1.7 & 43.60 & X21207\_112 & BLAGN & 0.720 & 283502201418119 & 20.04 & 19.12 & 18.60\\
  XWAS J000026.0-250649 & 2.0 & 43.13 & X21207\_058 & BLAGN & 0.433 & 283502201407236 & 21.10 & 20.58 & 19.28\\
  XWAS J000029.8-251213 & 7.1 & 42.70 & X21207\_029 &   ALG & 0.160 & 283502201402660 & 18.72 & 17.39 & 17.11\\
  XWAS J000031.8-245501 & 10.2 & 43.42 & X21207\_110 & NELG & 0.286 & 283502201417551 & 20.12 &99.99&99.99\\
  XWAS J000033.1-250917 & 0.2 & 41.87 & X21207\_046 & NELG & 0.324 & 283502201405124 & 20.33 & 19.01 &99.99\\
  XWAS J000034.6-250620 & 3.0 & 44.42 & X21207\_060 & BLAGN & 1.232 & 283502201407669 & 21.51 & 19.80 &99.99\\
  XWAS J000036.6-250104 & 1.5 & 43.34 & X21207\_092 & NELG & 0.592 & 564977178154666 &99.99& 21.09 &99.99\\
  XWAS J000100.1-250501 & 9.5 & 44.52 & X21207\_069 & BLAGN & 0.851 & 283781374297752 & 20.77 &99.99&99.99\\
  XWAS J000102.4-245849 & 9.5 & 43.81 & X21207\_101 & BLAGN & 0.433 & 283781374303399 & 22.28 & 20.13 &99.99\\
  XWAS J000106.7-250847 & 3.3 & 44.57 & X21207\_048 & BLAGN & 1.355 & 283781374294122 & 22.84 & 20.50 &99.99\\
  XWAS J000108.9-251613 & 0.6 & 44.62 & X21207\_016 & BLAGN & 2.800 & 283502201399377 & 21.98 &99.99&99.99\\
  XWAS J000109.2-245618 & 1.1 &99.99& X21207\_108 & star &0.000& 283502201416365 & 19.04 & 18.22 & 18.40\\
  XWAS J000122.6-250020 & 15.3 & 44.86 & X21207\_096 & BLAGN & 0.960 & 283502201412632 & 19.41 & 18.61 & 18.94\\
  XWAS J000222.4-255657 & 2.7 & 44.03 & X00821\_159 & BLAGN & 0.904 & 283502201366337 & 21.07 & 21.28 &99.99\\
  XWAS J000241.9-255414 & 0.6 & 41.21 & X00821\_176 &   ALG & 0.100 & 283502201368671 & 16.74 & 15.72 & 15.35\\
  XWAS J000253.5-260345 & 82.2 & 44.44 & X00821\_088 & BLAGN & 0.320 & 283502201360456 & 17.62 & 17.47 & 16.88\\
\hline\end{tabular}\\
\small{$^{(1)}$Absorbed X-ray flux in the 0.5--4.5\,keV energy band, in units of $10^{-14}$\,\fluxunit.}\\
\small{$^{(2)}$Logarithm of the absorbed X-ray luminosity in the 0.5--4.5\,keV energy band in \lumunit.}
\end{table*}

\begin{acknowledgements}
XMM-Newton project is an ESA science mission with instruments and contributions directly funded by ESA member states and NASA. This project is based on data obtained with the Anglo Australian Telescope's 2dF multi-fibre spectrograph. PE and AAH acknowledge support from the Spanish Plan Nacional de Astronom\'ia y Astrof\'isica under grant AYA2009-05705-E. PE, MP, SM, MW and JAT acknowledge support from the UK STFC research council. This work has been supported in part by the German DLR under contract numbers 50 OR 0404 and 50 OX 0201. The research leading to these results has received funding from the European Community's Seventh Framework Programme (/FP7/2007-2013/) under grant agreement No 229517. MK thanks for the support by the Deutsches Zentrum f\"ur Luft- und Raumfahrt (DLR) GmbH under contract No. FKZ 50 OR 0404. The Space Research Organisation of The Netherlands is supported financially by NWO, the Netherlands Organisation for Scientific Research. AC, RDC and PS acknowledge financial support from ASI (grant n. I/088/06/0, COFIS contract and grant n. I 009/10/0).
\end{acknowledgements}

\appendix
\section{Discrepancies with NED}\label{app:app1}
A detailed literature search provided a source characterisation (i.e., optical spectral classification and redshift) for 225 XWAS candidate counterparts. 
We compared 2dF identified sources with those previously classified according to NED. They are considered the same object if both detections are located within 2.5~arcsec and the published redshift is the same ($\pm$0.01) than that derived in our analysis. The majority of the NED classifications agree with ours except for a few exceptions, which are presented in Table~\ref{tab:ned}.

\begin{table*}

\caption{Discrepancies with NED}
\label{tab:ned}

\centering
\begin{tabular}{|c|c|c|c|c|c|c|}
\hline
  \multirow{2}{*}{Opt ID} &   {Type} &   {Redshift} &   \multirow{2}{*}{Source name} & {Type} & {Offset} & \multirow{2}{*}{Reference}\\
& XWAS & XWAS & & NED & {(arcsec)} &\\
\hline
  X20239\_073 & Gal & 0.113 & FCSS J033851.5-352650 & AGN1 & 0.6  & \citet{DellaCeca2004} \\
  X21207\_110 & NELG & 0.286 & HELLAS2XMM J000031.7-245459 & AGN1 & 2.1 & \citet{Fiore2003} \\
  X21223\_00064 & NELG & 0.275 & XLSS J022202.7-050942 & BLAGN & 1.3 & \citet{Garcet2007} \\
  X21226\_00115 & BLAGN & 0.149 & 2MASX J02255886-0500542 & NELG & 1.3 & \citet{Garcet2007} \\
  X21227\_00028 & BLAGN & 0.149 & APMUKS(BJ) B022438.41-051753.2 & NELG & 1.7 & \citet{Garcet2007} \\
  X21512\_00073 & BLAGN & 0.215 & XLSS J022253.5-042927 & NELG & 0.9 & \citet{Garcet2007} \\
  X21565\_00080 & BLAGN & 0.327 & XMDS J022649.0-042745 & NELG & 0.8 & \citet{Tajer2007} \\
  X21565\_00116 & NELG & 0.053 & 2MASX J02270078-0420209 & Sy1 & 0.0 & \citet{Lacy2007} \\
\hline\end{tabular}

\end{table*}

\bibliographystyle{aa}
\bibliography{bibliography}

\end{document}